\documentstyle[epsf]{article}

\def\abstract#1{\vskip 7mm 
        \begin{center}{\large Abstract}\par \smallskip
                \begin{minipage}[c]{12cm}
                        \small #1
                \end{minipage}
        \end{center}
}
\def\title#1{\begin{center}{\Large\bf #1}\end{center}}
\def\author#1{\vskip 5mm \begin{center}{#1}\end{center}}
\def\address#1{\begin{center}{\it #1}\end{center}}

\makeatletter
\@ifundefined{lesssim}{\def\lesssim{\mathrel{\mathpalette\vereq<}}}{}
\@ifundefined{gtrsim}{}{}
\def\vereq#1#2{\lower3pt\vbox{\baselineskip1.5pt \lineskip1.5pt
\ialign{$\m@th#1\hfill##\hfil$\crcr#2\crcr\sim\crcr}}}
\makeatother

\begin{document}

\begin{flushright}    
YITP-02-57 \\
OCU-PHYS-190 \\
AP-GR-6 \\
\end{flushright}      
\vskip 0.5cm

\title{Thick Domain Walls around a Black Hole}
\author{  
Yoshiyuki Morisawa,\footnote{E-mail:morisawa@sci.osaka-cu.ac.jp}
Daisuke Ida,\footnote{E-mail:d.ida@th.phys.titech.ac.jp}
Akihiro Ishibashi,\footnote{E-mail:akihiro@yukawa.kyoto-u.ac.jp}
 and 
Ken-ichi Nakao\footnote{E-mail:knakao@sci.osaka-cu.ac.jp}
        } 
\vskip5mm
\address{${}^{1,4}$\
Department of Physics, Osaka City University,
Osaka 558--8585, Japan,
         }
\address{${}^{2}$\
Department of Physics, Tokyo Institute of Technology,
Tokyo 152--8551, Japan,
        }
\address{${}^{3}$\
Yukawa Institute for Theoretical Physics, Kyoto University, 
Kyoto 606--8502, Japan \\
and \\
Enrico Fermi Institute, University of Chicago, 
Chicago, IL~60637, USA \\ 
        }
\abstract{
We discuss the gravitationally interacting system of a thick domain
wall and a black hole.
We numerically solve the scalar field equation in the Schwarzschild
spacetime and 
obtain a sequence of static axi-symmetric solutions
representing thick domain walls.
We find that, for the walls near the horizon, the Nambu--Goto approximation 
is no longer valid.  
}

\section{Introduction}
Topological defects are relics of cosmological phase transitions 
and their evolution is considered to have played 
an important role in cosmology (see {\it e.g.},~\cite{VilenkinBook}). 
They have so far attracted attention as, for example, 
a potential source of the cosmic structures, or as triggers of varieties 
of inflation~\cite{Vilenkin:1994pv,Bucher:1994gb,Yamamoto:1995sw}. 
Recently renewed interests especially in domain walls have been raised by 
considering them as a candidate for some kind of dark 
matter~\cite{Battye:1999eq}, or 
as a brane universe in higher dimensional theories 
(see {\it e.g.},~\cite{RS}). 
The topological defects have therefore been an important subject of 
recent study.

Domain walls and cosmic strings are of special interests in general relativity 
because they are extended objects with large tension and 
hence cause non-trivial gravitational 
effects~\cite{Vilenkin:1983hy,Ipser:1984db}. 
Then, it is intriguing to study how such extended relativistic objects  
interact with other extended objects, or a strong gravitational source 
like a black hole. 
In particular, domain walls or strings around a black hole may have 
experienced large deformation and be a possible source of gravitational waves. 
Such an expectation prompts us to study the gravitationally interacting 
system of topological defects and black holes.

To study configurations of walls and strings in a curved background, 
it is convenient to treat them as infinitely thin non-gravitating 
membranes whose dynamics obeys the Nambu--Goto action. 
So far, a number of works on defects--black-hole system have been done 
by using this membrane (thin-wall or -string) approximation. 
For example, the scattering problem of a Nambu--Goto string 
by a background black hole has been studied 
in detail~\cite{DeVilliers:1998nk,DeVilliers:1998xz,
Page:1998ya,DeVilliers:1999nm,Page:1999fd}. 
Christensen, Frolov, and Larsen~\cite{Christensen:1998hg,Frolov:1999td} 
considered Nambu--Goto walls embedded in the Schwarzschild black hole 
spacetime and found the static axisymmetric solutions. 
For the case where a thin wall is located at the equatorial plane of the
Reissner--Nordstr{\o}m--de Sitter black hole, the stability against
the perturbation is investigated by Higaki, Ishibashi, and
Ida~\cite{Higaki:2000yv}.
They showed that such a thin wall is unstable except the Schwarzschild
case. Motivated by brane-world scenario, Emparan, Horowitz, 
and Myers~\cite{Emparan:1999wa,Emparan:1999fd} constructed the
spacetime in which gravitating thin domain wall intersects a black
hole.

We should however notice that the thin-wall approximation is not always valid 
in a defects--black-hole system; there are some cases in which 
the thickness $w$ of defect becomes comparable with or 
even exceeds the size of a black hole $R_g$. 
One of such examples is given by considering primordial black holes 
which evaporate at the present 
epoch~(see {\it e.g.}~\cite{Cline:1998ft,Carr:YKIS99,Yokoyama:YKIS99}) 
and defects formed during a late time phase transition 
at $\lesssim 100\mbox{\rm MeV}$~\cite{Hill:1989vm,Battye:1999eq}, 
as discussed in our previous work~\cite{Morisawa:2000qq}.

In this paper, we shall treat a thick domain wall which 
gravitationally interacts with a black hole. 
This work is complementary to our previous work~\cite{Morisawa:2000qq}, 
in which we have shown that the thick domain wall intersecting 
a black hole can exist as a static configuration of a scalar field 
for the case where the core of the wall is located at the equatorial 
plane of the black hole. 
This wall can be described by the world sheet of a Nambu--Goto
membrane since the equatorial plane is a minimal
surface~\cite{Christensen:1998hg,Frolov:1999td}. 
In the present work, we remove the limitation that the core of wall is  
located at the equatorial plane of the black hole, aiming at examining 
the validity of Nambu--Goto approximation in the black hole spacetime. 

For thorough study of the wall-black hole system, it is necessary to take 
the gravitational back reaction of the wall into consideration. 
In the present paper, we shall however focus on a non-gravitating 
domain wall as a tractable case. 
This test wall assumption might be valid when the symmetry braking  
scale of the scalar field is much lower than Planck scale 
as shown in~\cite{Morisawa:2000qq} by dimensional analysis. 
For the case that a domain wall is on the equatorial plane of 
a black hole,
Emparan, Gregory, and Santos~\cite{Emparan:2000fn}
and~Rogatko~\cite{Rogatko:2001vv}
recently studied a system of thick domain walls with a black hole 
sitting on it, taking the gravitational backreaction into account. 
They also discussed the quantum nucleation of such a wall-black hole system. 

In the next section, we derive the basic equation and discuss the boundary
conditions which represent the situation we want to study. 
Our setup and the basic equations are essentially 
the same as those in the previous work~\cite{Morisawa:2000qq}, 
apart from the boundary conditions. 
In section~\ref{sec:results}, we show the numerical results.
In section~\ref{sec:comparison}, we analyze the numerical
solutions, and discuss the validity of Nambu--Goto approximation.
We summarize our work in section~\ref{sec:summary}.
Throughout this paper, we use units such that $c=\hbar=G=1$ unless otherwise 
stated.

\section{The basic equation and the boundary conditions}
\label{sec:equation}

We shall consider a static thick domain wall constructed by a scalar field 
with self-interaction in the Schwarzshild black hole spacetime. 
The metric of the background Schwarzschild black hole is written in terms of 
the isotropic coordinates $\{t,r,\vartheta,\varphi\}$ as 
\begin{equation}
g=
  -\left(\frac{2r-M}{2r+M}\right)^2{\rm d}t^2
  +\left(1+\frac{M}{2r}\right)^4
  \left[
        {\rm d}r^2+r^2({\rm d}\vartheta^2+\sin^2\vartheta {\rm d}\varphi^2) 
  \right]. 
\end{equation} 
We are concerned with the region outside the event horizon, 
where $r \ge M/2$.

Let us consider a real scalar field $\phi$ with a potential $V[\phi]$, 
of which Lagrangian is given by 
\begin{equation}
{\cal L}=
  -\frac{1}{2}(\nabla^\mu \phi)( \nabla_\mu \phi) - V[\phi] . 
\label{eq:Lag}
\end{equation}
The equation of motion for $\phi$ is
\begin{equation}
\nabla^2\phi-\frac{\partial V}{\partial \phi}=0.
\label{eq:EOM}
\end{equation}
In this paper, we consider a familiar type of potential which has a
discrete set of degenerate minima; the $\phi^4$ potential
\begin{equation}
V[\phi]=\frac{\lambda}{4}(\phi^2-\eta^2)^2.
\end{equation}
Note that for this potential, Eq.~(\ref{eq:EOM}) has an analytic 
solution 
\begin{equation}
\phi(z)=\eta\tanh\left[\sqrt{\frac{\lambda}{2}}\eta (z-z_c)\right]
\label{eq:flat4}
\end{equation}
in the flat spacetime $g=-{\rm d}t^2+{\rm d}x^2+{\rm d}y^2+{\rm d}z^2$. 
This solution represents a static and plane-symmetric domain wall, 
and is characterized by the thickness of the wall
\begin{equation}
w = \frac{1}{\sqrt{\lambda}\eta},
\end{equation}
and the position  $z_c$ of the wall's core. 

We will search the static and axi-symmetric solutions 
$\phi=\phi(r,\vartheta)$ which represent domain walls
in the Schwarzschild background.
Let us introduce a dimensionless parameter 
\begin{equation}
\epsilon= {M \over 2w} ,
\end{equation} 
and dimensionless variables 
\begin{equation}
\rho= 2r M^{-1},~~~\Phi(\rho,\vartheta)= \eta^{-1} \phi(r,\vartheta) .
\end{equation}
The parameter $\epsilon$ is just a ratio of the horizon radius to the
wall's thickness, namely if $\epsilon$
is smaller (larger) than unity, then the wall is said to be thick
(thin) as compared to the size of the black hole.
In terms of these variables, 
the equation of motion (\ref{eq:EOM}) is written as 
\begin{equation}
\left(\frac{\rho}{\rho+1}\right)^4
\left[
\frac{\partial^2}{\partial \rho^2}
+\frac{2\rho}{(\rho^2-1)}\frac{\partial}{\partial \rho}
+\frac{1}{\rho^2}\left(\frac{\partial^2}{\partial\vartheta^2}
+\cot\vartheta\frac{\partial}{\partial\vartheta}\right)
\right]\Phi
=
\epsilon^2\frac{\partial U}{\partial \Phi},
\label{eq:basic}
\end{equation}
where the dimensionless potential 
$U[\Phi]=V[\phi]/\lambda\eta^4$ is defined. 
$U$ has minima at $\Phi=\pm 1$. 
Since the equation (\ref{eq:basic}) is elliptic, the relaxation method
is useful to solve the discretized version of Eq.~(\ref{eq:basic}).

We now consider the boundary conditions suitable for our purpose.
First, the regularity of the scalar field at the symmetry axis is
given by the Neumann boundary conditions
\begin{equation}    
\left.\frac{\partial\Phi}{\partial\vartheta}\right|_{\vartheta=0}= 
\left.\frac{\partial\Phi}{\partial\vartheta}\right|_{\vartheta=\pi}=0.
\label{eq:BCaxis}
\end{equation}
Second, the regularity of the scalar field at the event horizon
$\{\rho=1\}$ is given by the Neumann boundary condition
\begin{equation}
\left.\frac{\partial\Phi}{\partial\rho}\right|_{\rho=1}=0.
\label{eq:BChorizon}
\end{equation}
This condition is a consequence of the requirement that $\Phi$ with
its first derivatives is regular at the horizon.
This then implies that the energy density observed by a freely falling
observer remains finite at the event horizon.
Finally, in practice, the domain of the numerical integration is
inevitably finite, so that we need an asymptotic boundary condition at
$\rho=\rho_{\rm max}$ for $\rho_{\rm max}\gg 1$. 
All the information about the position of the wall is controlled
by this boundary condition.
As this condition, in this paper, we adopt the condition explained 
in the following paragraph, which may describe an adiabatic capture 
process of a thick wall by a black hole.  

Since the background spacetime is asymptotically flat,
we can expect that there exist flat wall solutions far away from
the black hole and they are well approximated by analytic
solutions~(\ref{eq:flat4}).
We can therefore impose the Dirichlet boundary condition 
\begin{equation}
\Phi|_{\rho=\rho_{\rm max}} 
= \tanh [2^{-1/2}\epsilon(\rho_{\rm max}\cos\vartheta - z_c)], 
\label{eq:BCasymptotic1}
\end{equation}
for $|z_c|\gg 1$, where $z_c$ expresses how far the wall is away from the 
equatorial plane of the black hole. 
However, in general, imposing the condition~(\ref{eq:BCasymptotic1}) is not 
appropriate for the $|z_c|\approx 1$ case. 
In fact, no static axi-symmetric Nambu--Goto membrane whose
asymptotic surface is flat exists
unless the membrane is just lying on the equatorial
plane~\cite{Christensen:1998hg}.
Thus we shall adopt the following procedure. 
We first solve the equation of motion~(\ref{eq:basic}) 
under the boundary condition~(\ref{eq:BCasymptotic1}) with 
a sufficiently large $z_c$. 
Next, ``parallelly transporting'' the obtained solution
$\Phi_{(1)}(\rho,z;z_c)$ ($z:=\rho\cos\vartheta$) from $z$
to $z - \Delta z$ along the $z$-axis, we have a configuration
$\Phi_{(1)}(\rho,z+\Delta z;z_c)$, which is not necessarily a
solution to Eq.~(\ref{eq:basic}).
Provided $\Delta z$ is sufficiently small, 
the configuration $\Phi_{(1)}(\rho,z + \Delta z;z_c)$ is thought to well 
approximate a new wall solution $\Phi_{(2)}(\rho,z;z_c)$ which is closer
to the black hole than $\Phi_{(1)}(\rho,z;z_c)$.
Hence we can use $\Phi_{(1)}(\rho,z + \Delta z;z_c)$ as the initial value for 
the relaxation method to obtain the solution $\Phi_{(2)}(\rho,z;z_c)$. 
Note that the value of $\Phi_{(1)}(\rho,z + \Delta z;z_c)$ at the boundary of 
the computation region also gives the Dirichlet boundary condition 
for obtaining $\Phi_{(2)}(\rho,z;z_c)$. 
Then, the parallelly transported configuration $\Phi_{(2)}(\rho,z + \Delta z;z_c)$ 
gives the initial value and the boundary value 
for obtaining a next solution $\Phi_{(3)}(\rho,z;z_c)$. 
By repeating this procedure, we can obtain a sequence of the wall solutions
$\Phi_{(n)}(\rho,z;z_c)$, some of which will be very close to 
the equatorial plane.

\section{Numerical results}
\label{sec:results}

\begin{figure}
\centerline{
\epsfxsize=14.0cm
\epsfbox{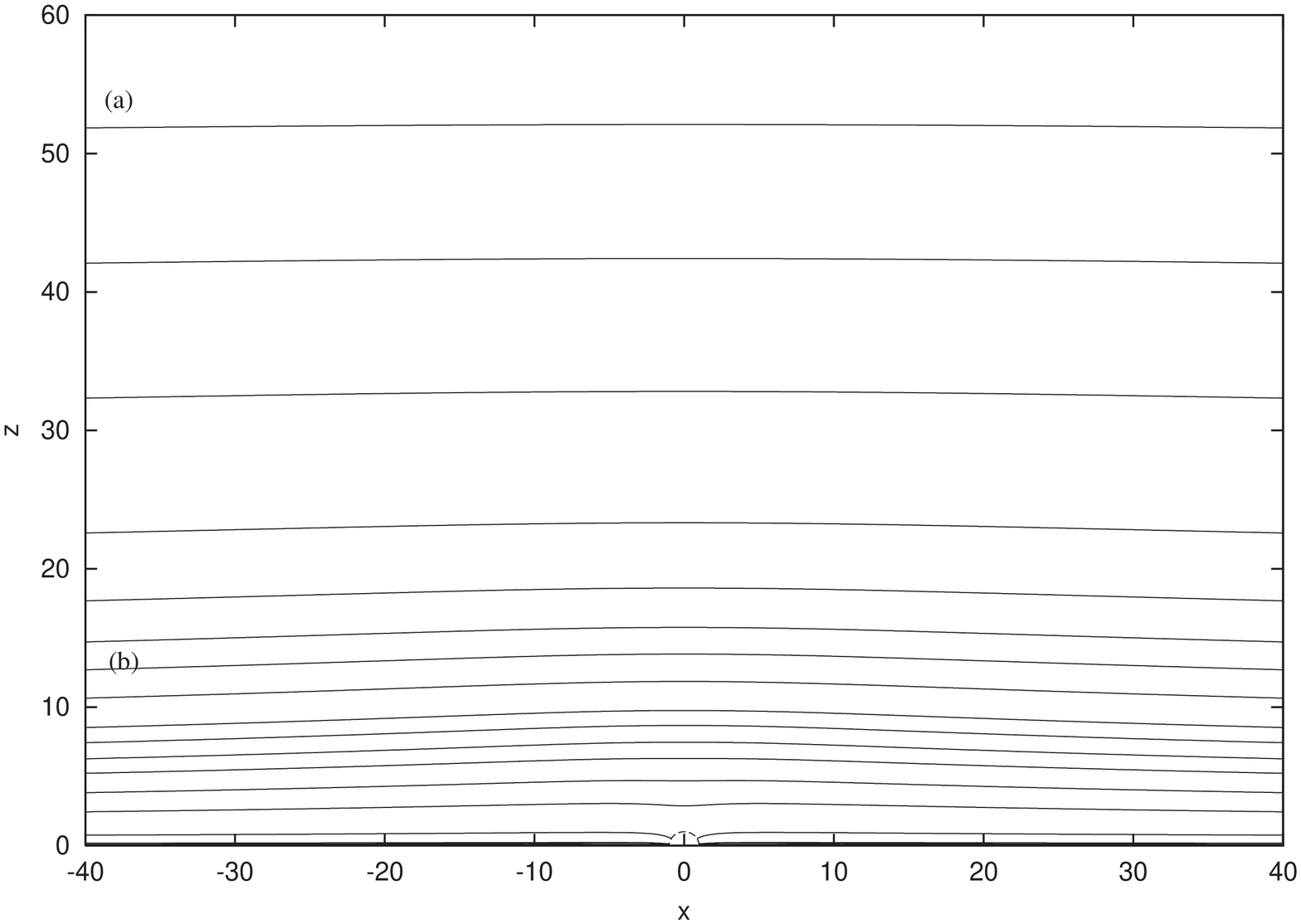}
}
\centerline{
\epsfxsize=14.0cm
\epsfbox{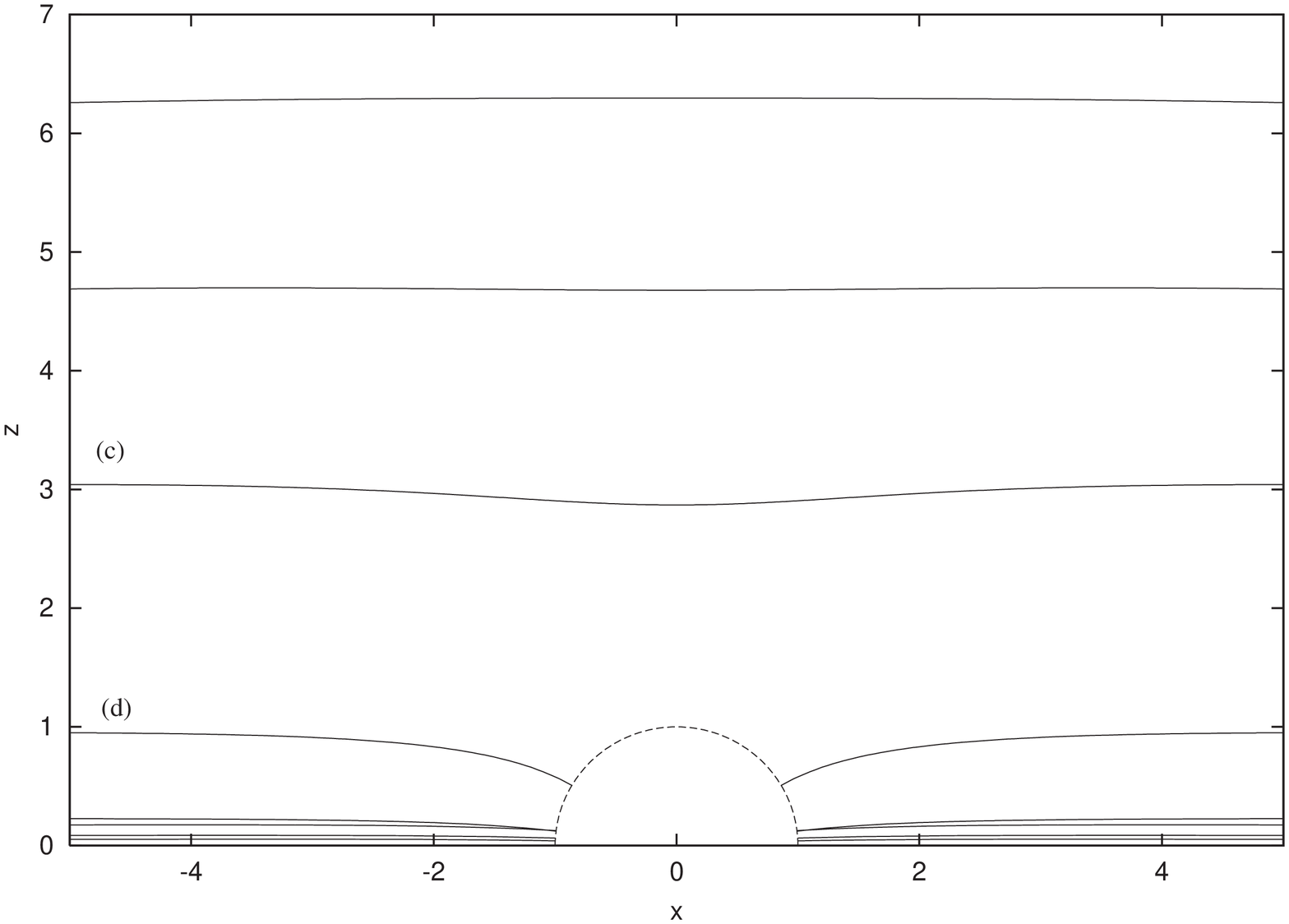}
}
\caption{The sequence of the core surfaces (where $\Phi=0$) of the wall 
solutions whose thickness are an order of magnitude larger than the
Schwarzschild radius, {\it i.e.} $\epsilon=0.1$. 
Here the Cartesian coordinates 
$(x,z)=(\rho\sin\vartheta,\rho\cos\vartheta)$ are used. 
The curves must be rotated around $z$ axis to obtain the full spatial
geometry of the core surfaces. 
The dashed (semi-)circle at the origin denotes the event horizon.
Some of the walls are intersecting the horizon. 
The uppermost wall located about $z=50$ was calculated under the
boundary condition (\ref{eq:BCasymptotic1}), and starting from which 
we obtained the other solutions. 
The lower panel is close-up to the neighborhood of the event horizon.
The solutions correspond to (a),(b),(c) and (d) are shown in
Figs.~\ref{fig:kink_th10_far}, \ref{fig:kink_th10_middle},
\ref{fig:kink_th10_near} and \ref{fig:kink_th10_cross}, respectively.
}
\label{fig:coresurf_th10}
\end{figure}

\begin{figure}
\centerline{
\epsfxsize=14.0cm
\epsfbox{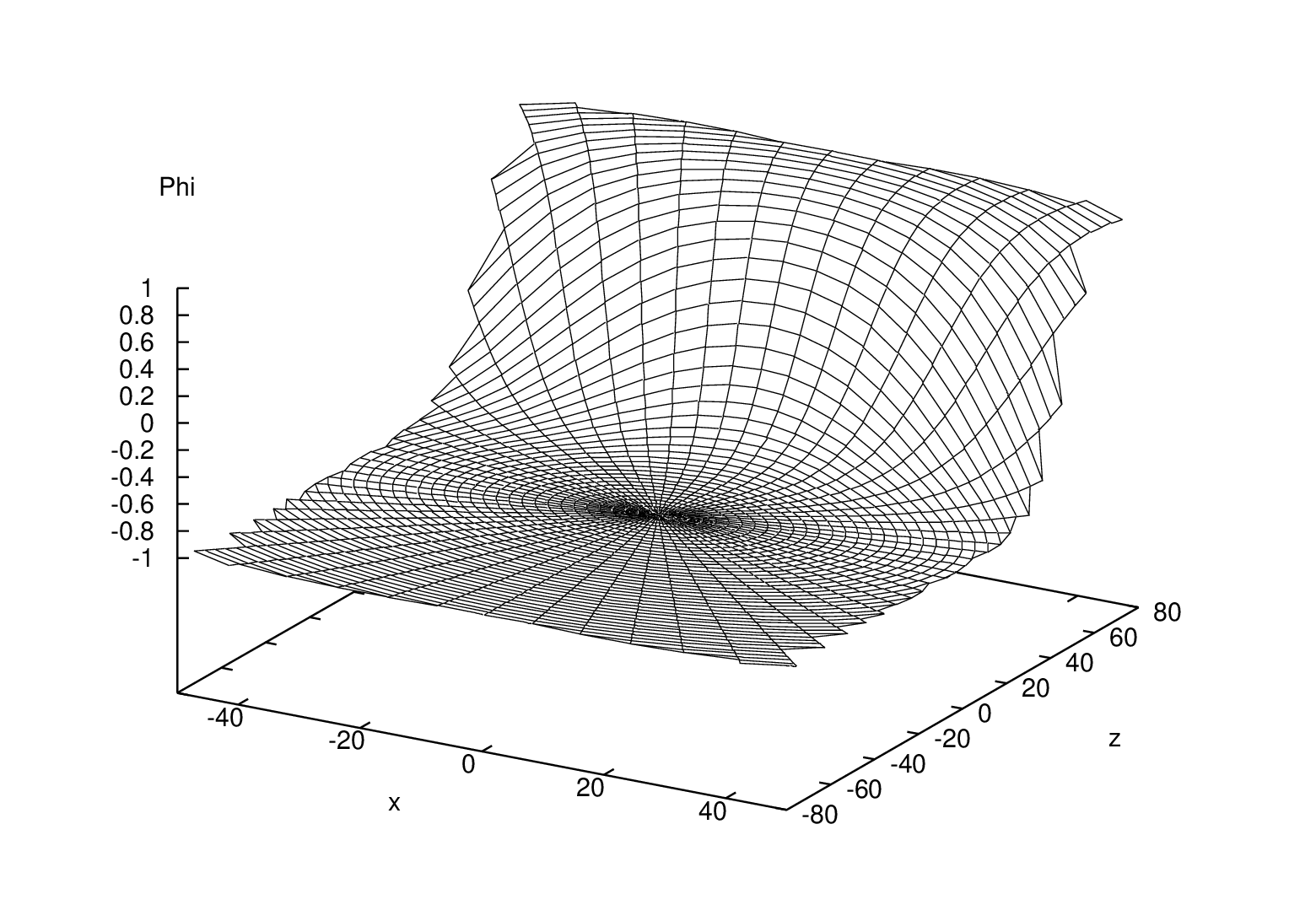}
}
\centerline{
\epsfxsize=14.0cm
\epsfbox{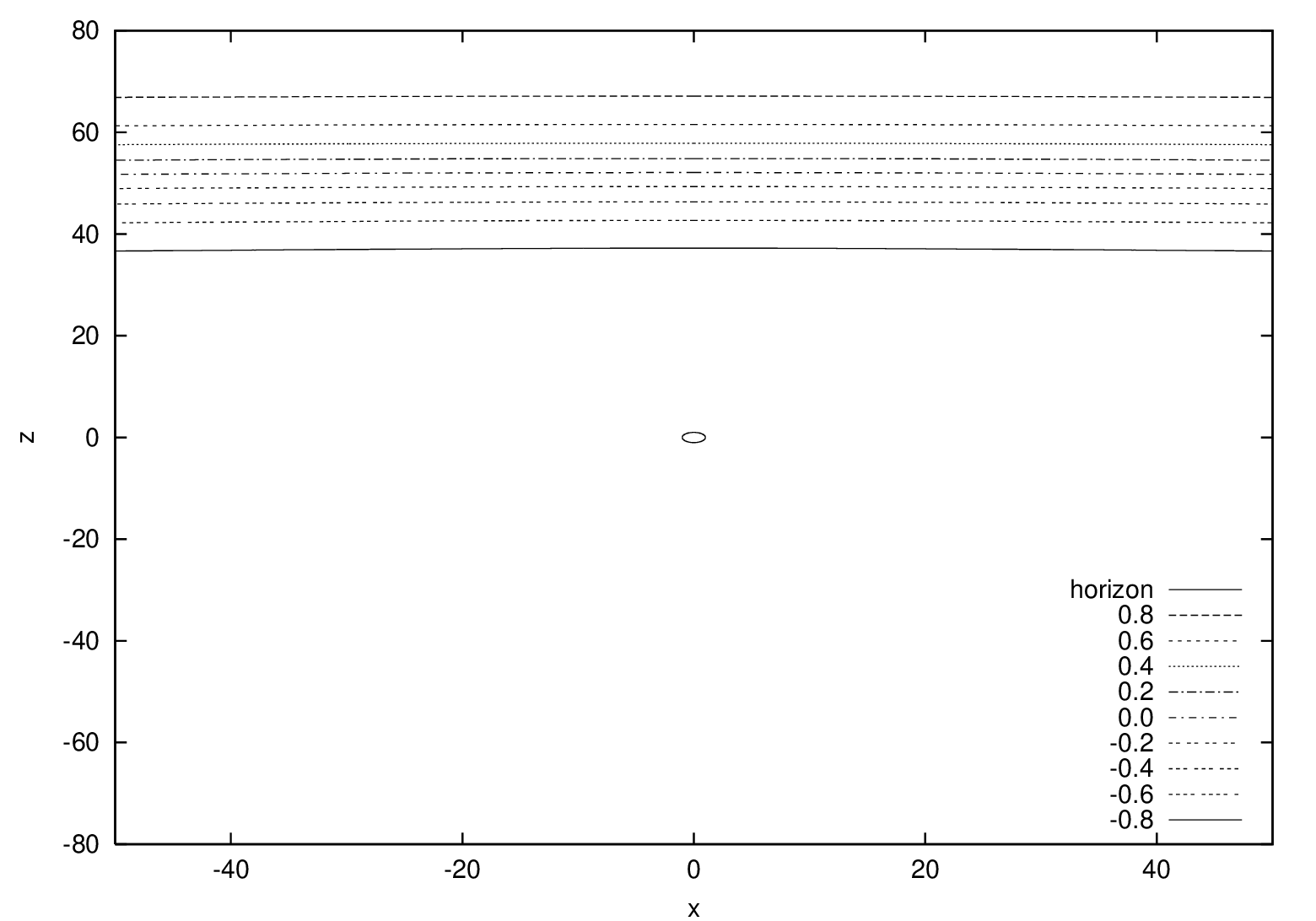}
}
\caption{(a) The wall solution $\Phi(x,z)$ obtained 
under the boundary condition~(\ref{eq:BCasymptotic1}) 
for $\epsilon=0.1$, $z_c=50$.
The upper panel shows the birds eye view of $\Phi(x,z)$.
It shows a kink structure localized around $z\sim50$ with the thickness
$w\sim10$.
The lower panel shows the contour plot of $\Phi(x,z)$.
Each line corresponds to
$\Phi=0.8,0.6,0.4,0.2,0.0,-0.2,-0.4,-0.6,-0.8$.
The circle $\rho=1$ is the horizon.
The contour lines are almost parallel to the equatorial plane.
We see that the almost flat domain wall arises far away from the black
hole.
}
\label{fig:kink_th10_far}
\end{figure}

\begin{figure}
\centerline{
\epsfxsize=14.0cm
\epsfbox{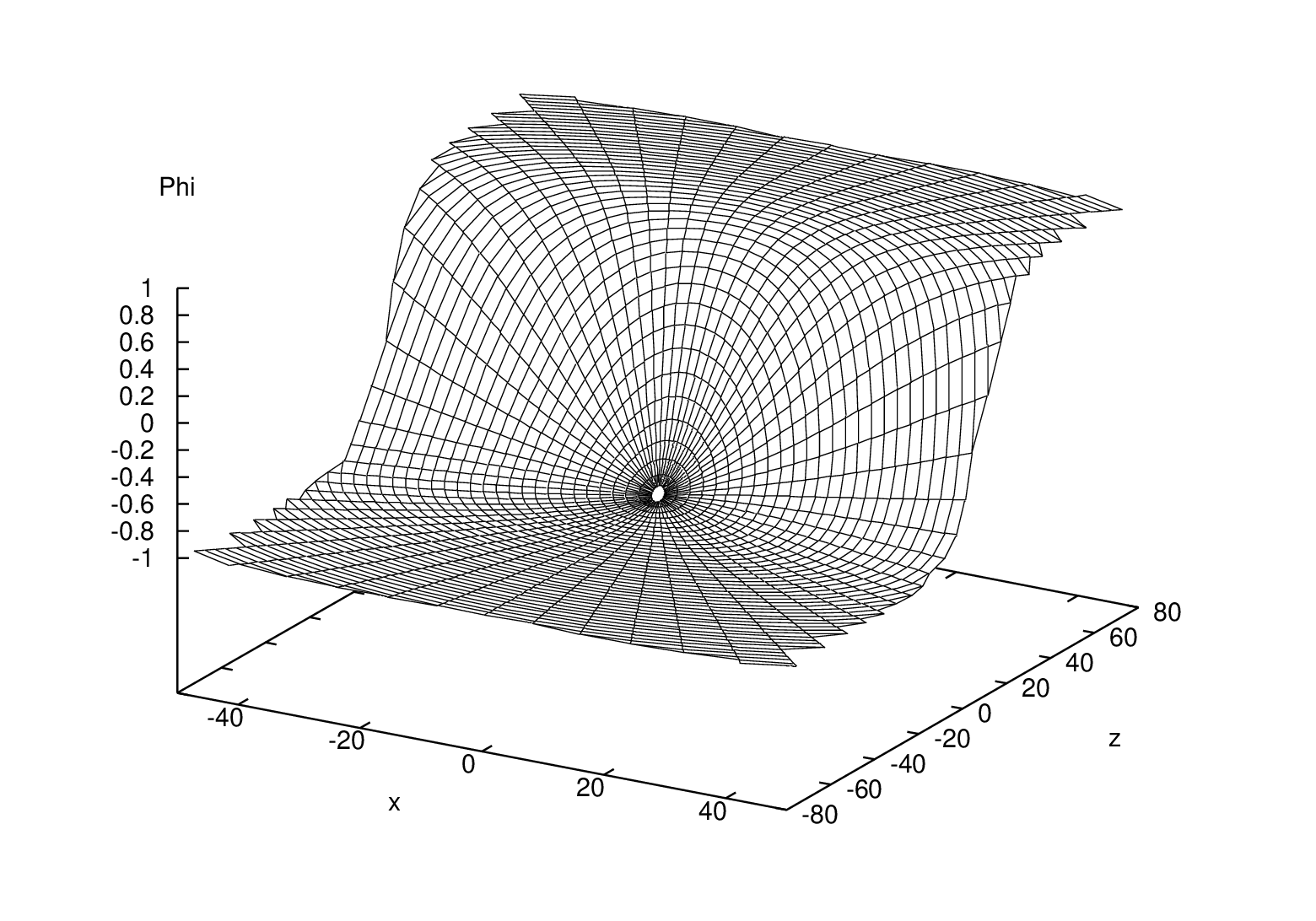}
}
\centerline{
\epsfxsize=14.0cm
\epsfbox{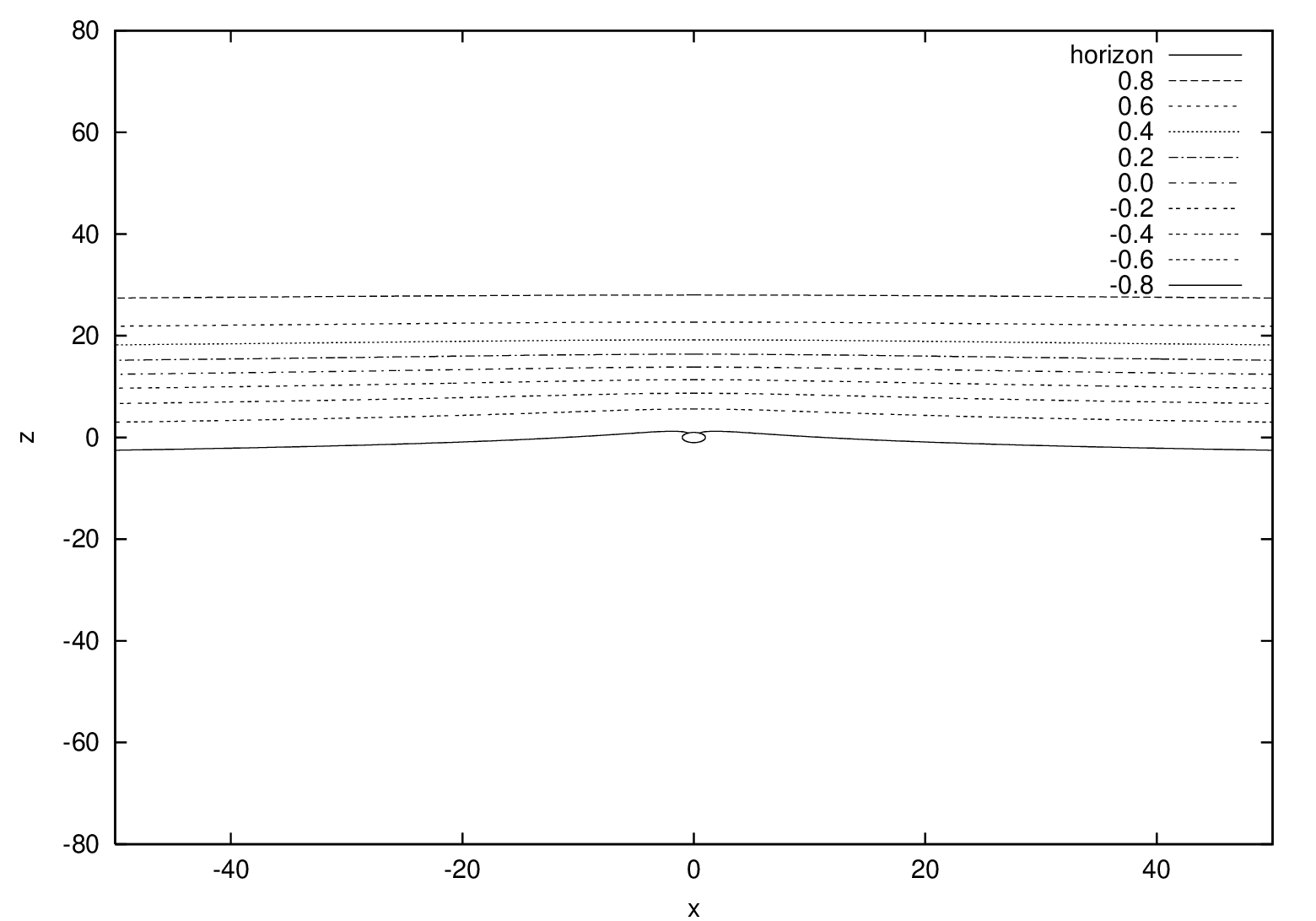}
}
\caption{(b) The wall solution $\Phi(x,z)$ whose core surface lies
about $z=10$.
The upper panel shows the birds eye view of $\Phi(x,z)$.
It shows a kink structure localized around $z\sim10$ with the thickness
$w\sim10$.
The separation from the equatorial plane of the black hole is comparable 
to the thickness of the wall.
The lower panel shows the contour plot of $\Phi(x,z)$.
Each line corresponds to
$\Phi=0.8,0.6,0.4,0.2,0.0,-0.2,-0.4,-0.6,-0.8$.
The circle $\rho=1$ is the horizon.
The contour lines are slightly bent like as the membrane solutions
in~\cite{Christensen:1998hg,Frolov:1999td}.
We see the bent wall near by the black hole.
}
\label{fig:kink_th10_middle}
\end{figure}

\begin{figure}
\centerline{
\epsfxsize=14.0cm
\epsfbox{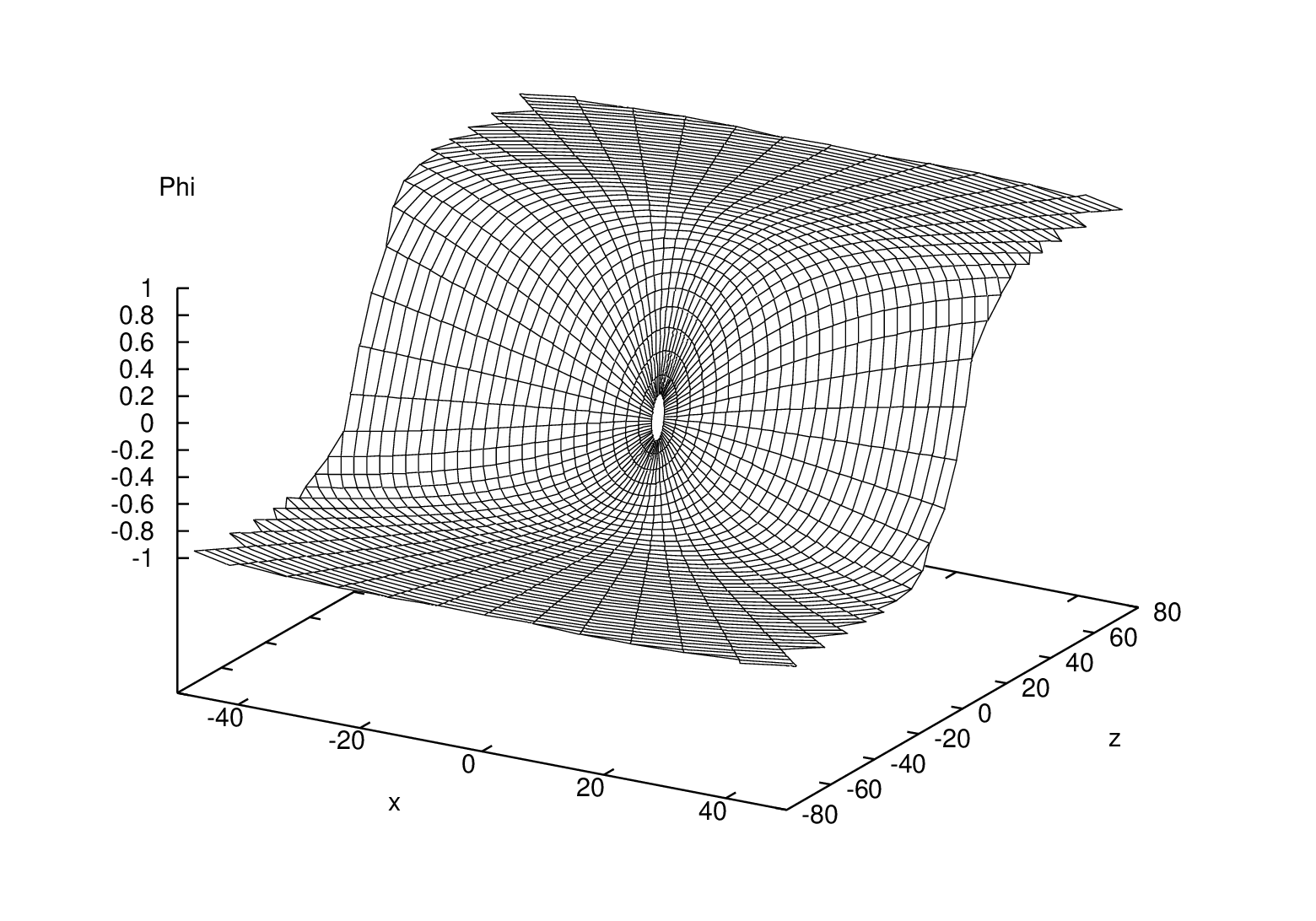}
}
\centerline{
\epsfxsize=14.0cm
\epsfbox{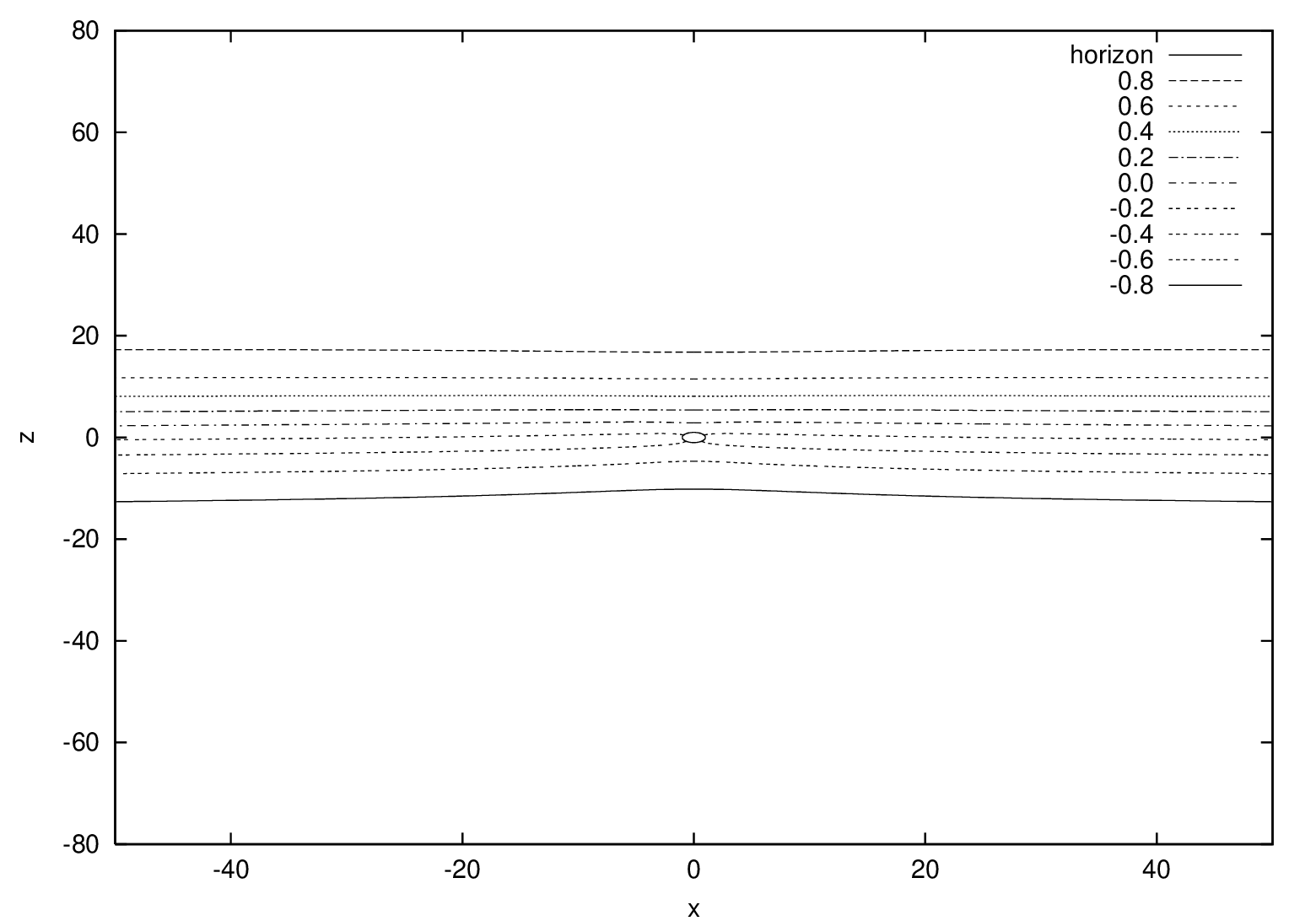}
}
\caption{(c) The wall solution $\Phi(x,z)$ whose core surface lies
near the black hole but does not intersect the horizon.
The upper panel shows the birds eye view of $\Phi(x,z)$.
It shows a kink structure localized around $z\sim1$ with the thickness
$w\sim10$.
The lower panel shows the contour plot of $\Phi(x,z)$.
Each line corresponds to
$\Phi=0.8,0.6,0.4,0.2,0.0,-0.2,-0.4,-0.6,-0.8$.
The circle $\rho=1$ is the horizon.
We see that the black hole is inside the thick wall.
}
\label{fig:kink_th10_near}
\end{figure}

\begin{figure}
\centerline{
\epsfxsize=14.0cm
\epsfbox{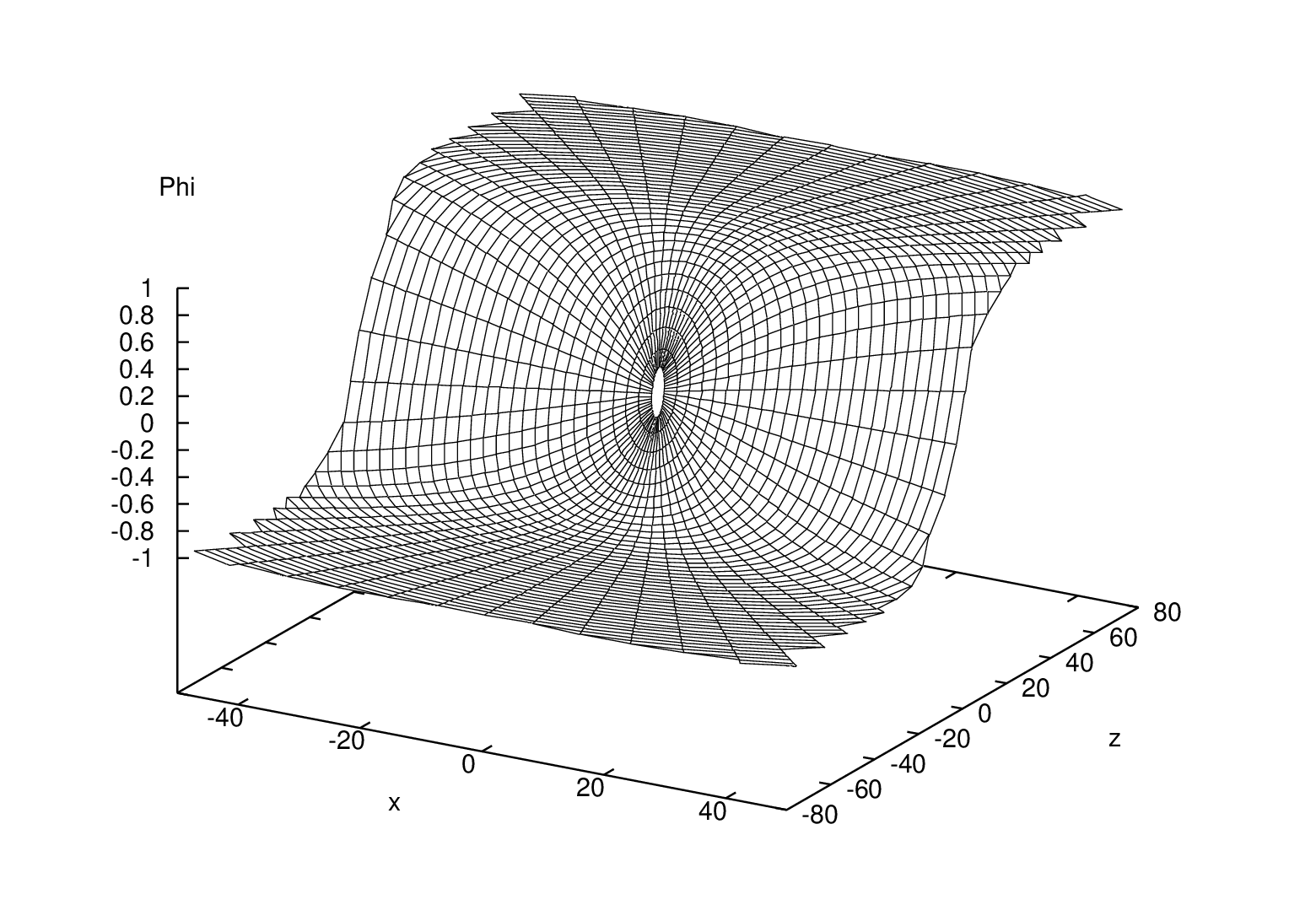}
}
\centerline{
\epsfxsize=14.0cm
\epsfbox{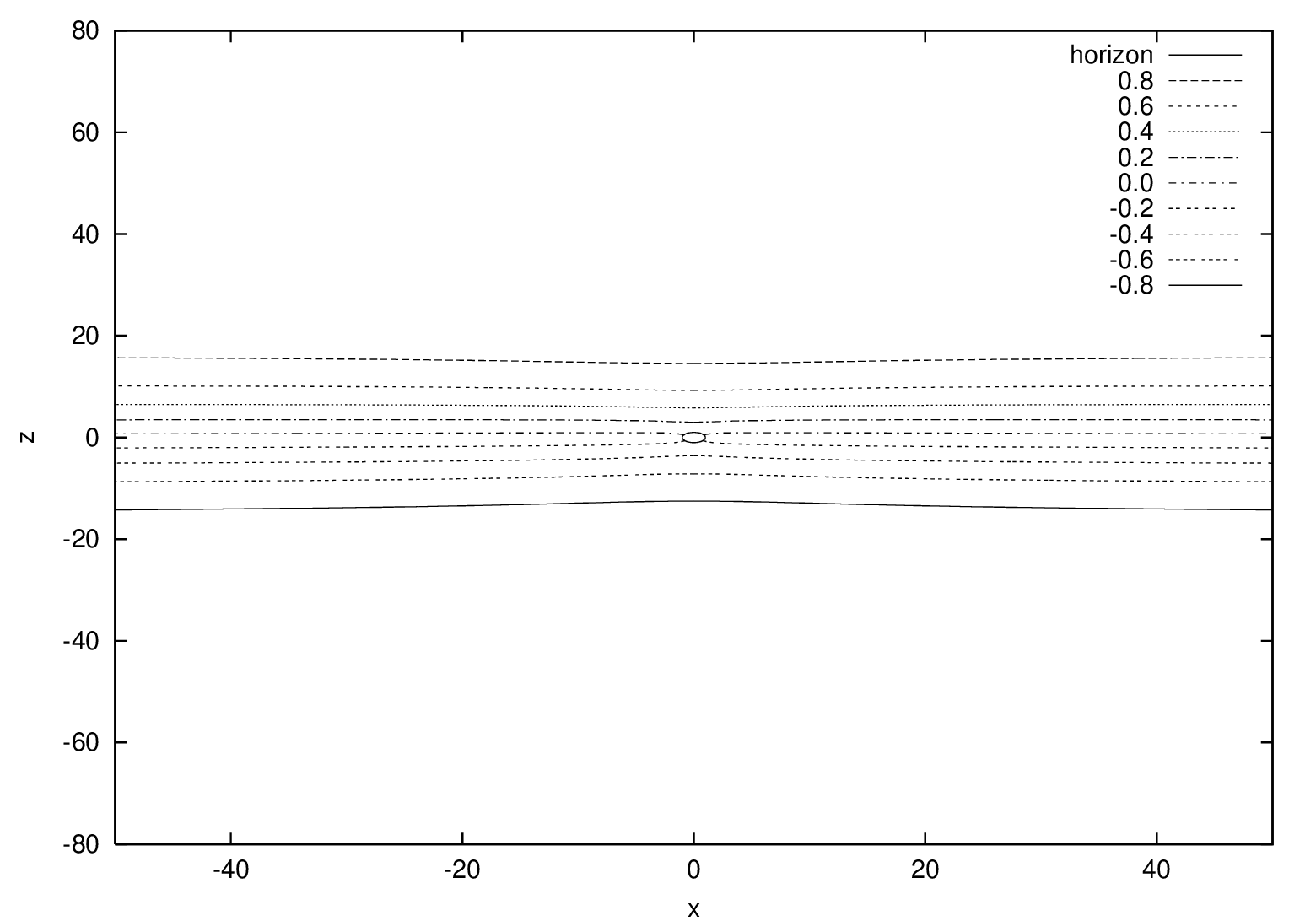}
}
\caption{(d) The wall solution $\Phi(x,z)$ whose core surface
intersects the horizon.
The upper panel shows the birds eye view of $\Phi(x,z)$.
It shows a kink structure around the equatorial plane with the thickness 
$w\sim10$.
The lower panel shows the contour plot of $\Phi(x,z)$.
Each line corresponds to
$\Phi=0.8,0.6,0.4,0.2,0.0,-0.2,-0.4,-0.6,-0.8$.
The circle $\rho=1$ is the horizon.
We see that the black hole is inside the thick wall.
}
\label{fig:kink_th10_cross}
\end{figure}

Following the procedure mentioned in the previous section,
we first solve Eq.~(\ref{eq:basic}) under the boundary 
conditions~(\ref{eq:BCaxis}),~(\ref{eq:BChorizon}), 
and~(\ref{eq:BCasymptotic1}) with $\epsilon=0.1$,  
$z_c = 50$. We obtain the solution $\Phi_{(1)}(\rho,z;50)$ 
using the relaxation method. 
Then, regarding $\Phi_{(1)}(\rho,z;50)$ as the initial
configuration for our subsequent computation, we solve
Eq.~(\ref{eq:basic}).
As a result, we obtain a sequence of wall configurations 
around the black hole. 
The core surfaces (where $\Phi=0$) of the obtained wall solutions are  plotted 
in Fig.~\ref{fig:coresurf_th10}.
We show the scalar field configurations $\Phi(x,z)$ 
for the following four typical cases: 
(a)~the wall solution $\Phi_{(1)}(\rho,z;50)$ far away from the black hole,  
(b)~the wall which is away from the black hole 
    at the distance comparable to the wall thickness, 
(c)~the wall whose core surface is located near the black hole 
     but does not intersect the horizon, 
and  
(d)~the wall whose core surface intersects the horizon, 
in Figs.~\ref{fig:kink_th10_far}--\ref{fig:kink_th10_cross}, respectively. 
In each figure,
the upper panel shows the birds eye view of $\Phi(x,z)$,
and the lower panel shows the contour plot of $\Phi(x,z)$.
Each contour line corresponds to
$\Phi=0.8$, $0.6$, $0.4$, $0.2$, $0.0$, $-0.2$, $-0.4$, $-0.6$, $-0.8$.

The numerical solution (a) has a kink structure localized around
$z\sim50$ with the thickness $w\sim10$.
Its contour lines are almost parallel to the equatorial plane.
We see that the almost flat domain wall arises far away from the black
hole
(Fig.~\ref{fig:kink_th10_far}).
The solution (b) has a kink structure localized around $z\sim10$
with the thickness $w\sim10$.
The separation from the equatorial plane of the black hole is comparable 
to the thickness of the wall.
Its contour lines are slightly bent but almost flat
(Fig.~\ref{fig:kink_th10_middle}).
The solution (c) and~(d) have kink structure localized around the
equatorial plane with the thickness $w\sim10$.
We see that the black hole is inside the thick wall
(Figs.~\ref{fig:kink_th10_near} and~\ref{fig:kink_th10_cross}).

\begin{figure}
\centerline{
\epsfxsize=14.0cm
\epsfbox{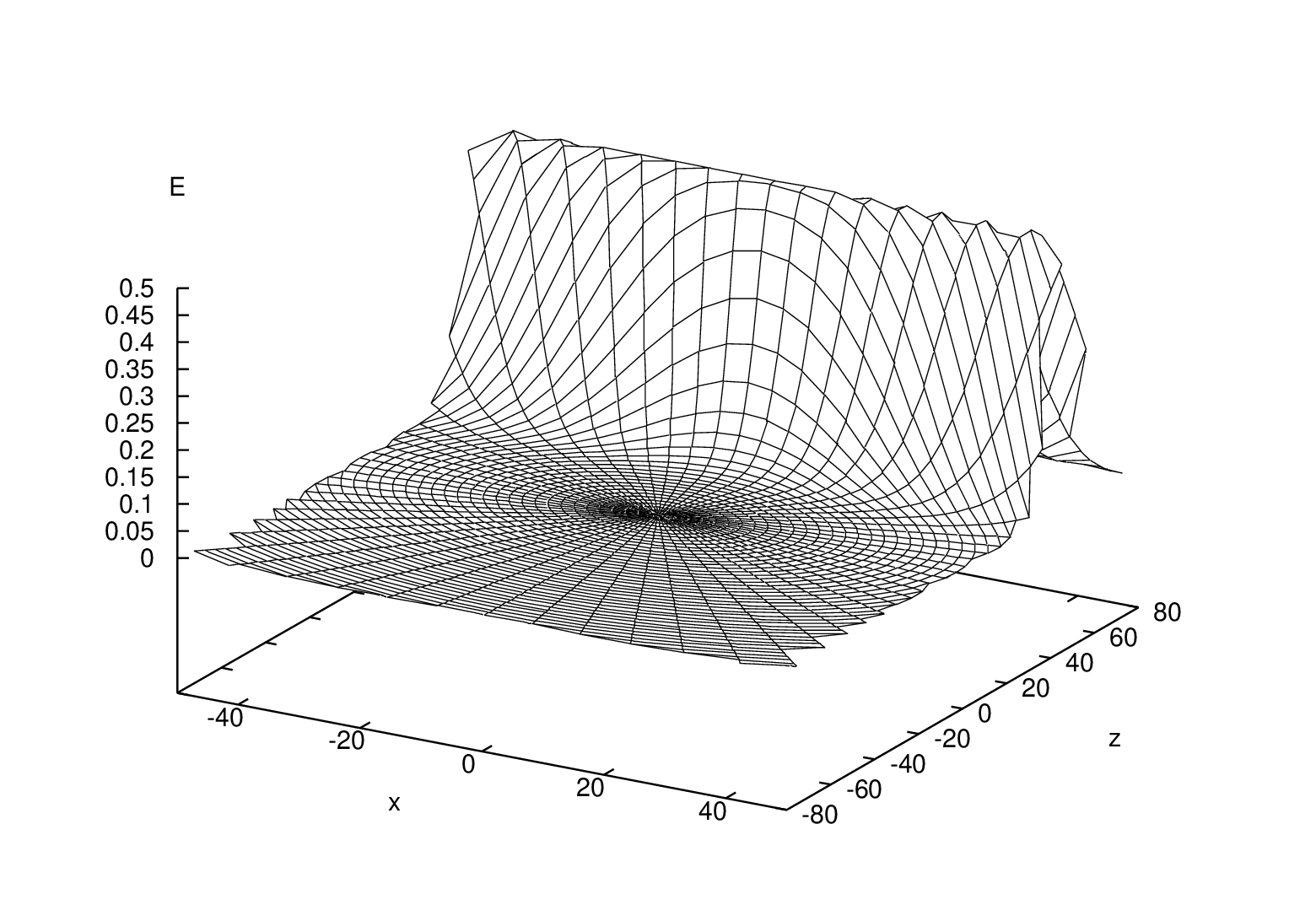}
}
\centerline{
\epsfxsize=14.0cm
\epsfbox{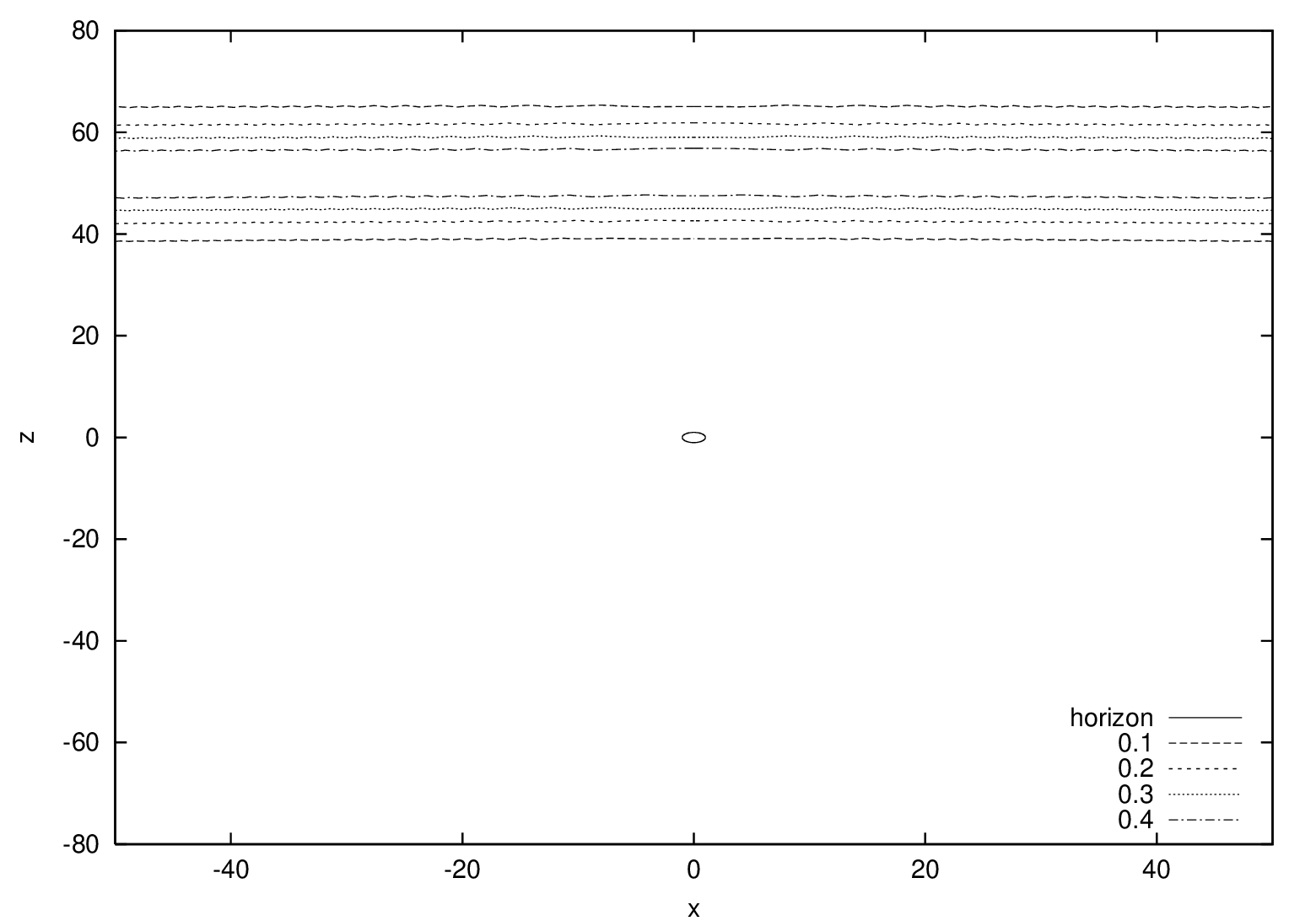}
}
\caption{(a) The energy density $E(x,z)$ of the wall solution shown in 
Fig.~\ref{fig:kink_th10_far}.
The upper panel shows the birds eye view of $E(x,z)$.
The almost flat domain wall with the thickness $w\sim10$ arises far
away from the black hole.
The lower panel shows the contour plot of $E(x,z)$.
Each line corresponds to $E=0.1,0.2,0.3,0.4$.
The circle $\rho=1$ is the horizon.
The energy density distribution is almost homogeneous along the wall.
}
\label{fig:wall_th10_far}
\end{figure}

\begin{figure}
\centerline{
\epsfxsize=14.0cm
\epsfbox{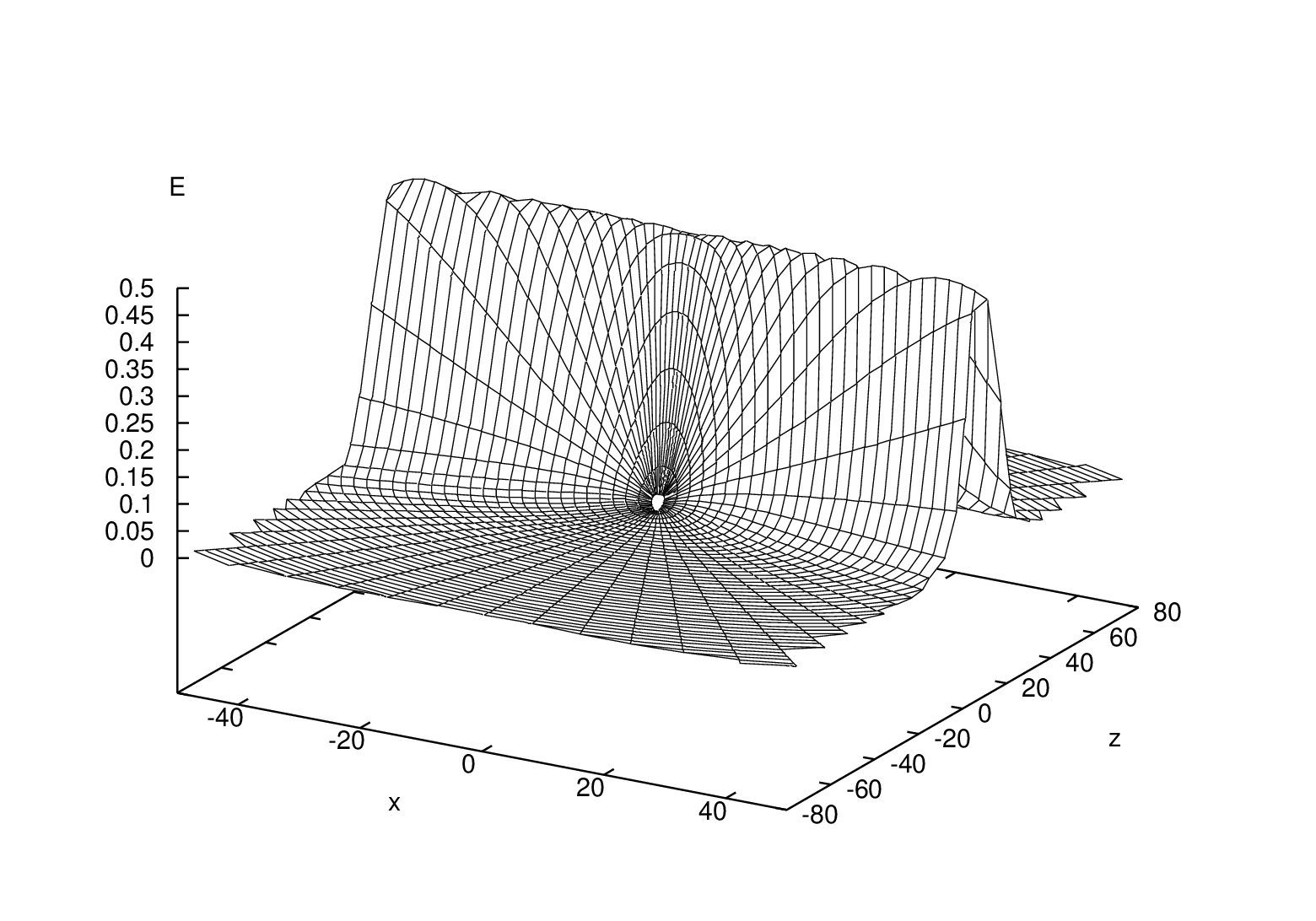}
}
\centerline{
\epsfxsize=14.0cm
\epsfbox{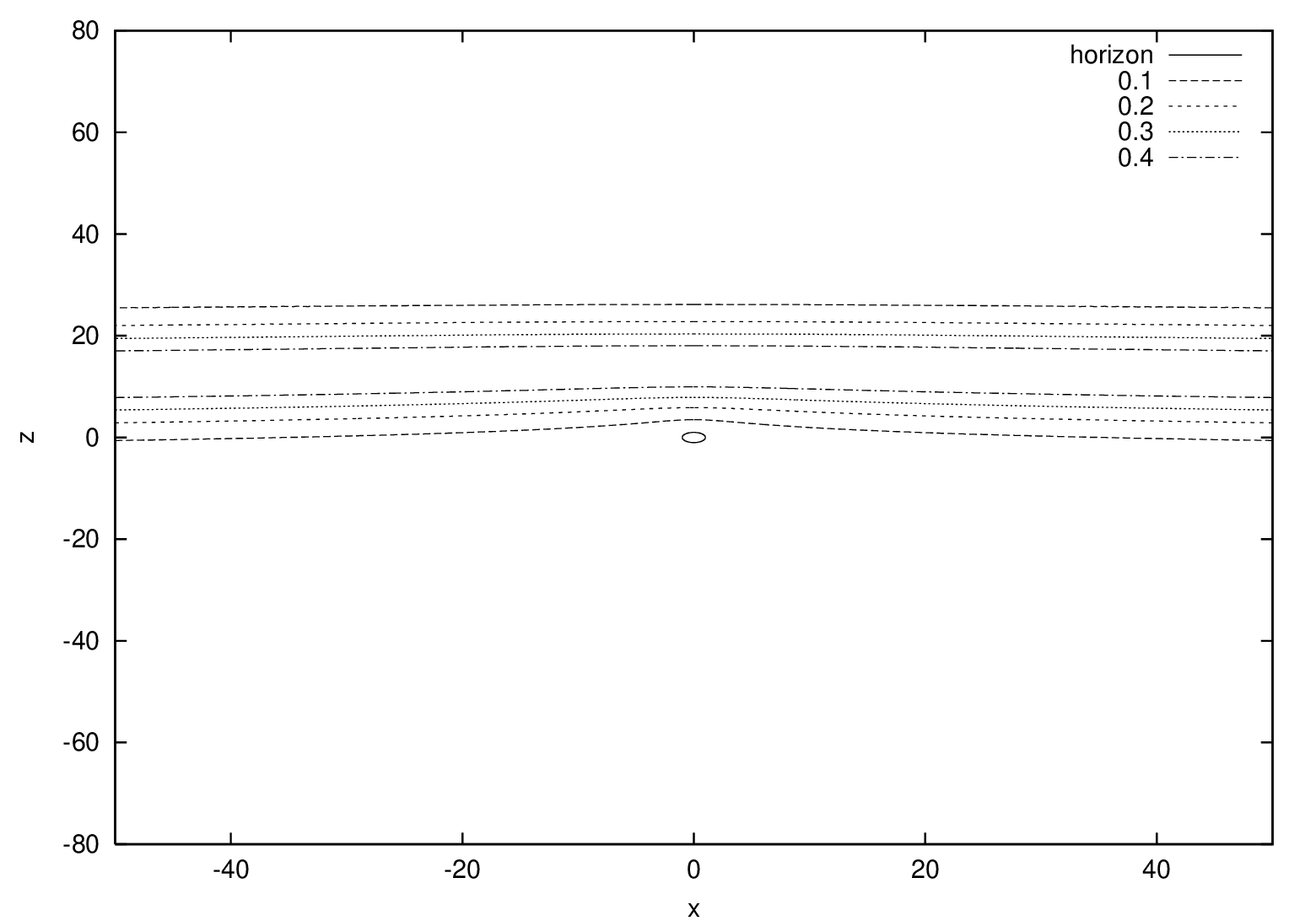}
}
\caption{(b) The energy density $E(x,z)$ of the wall solution in 
Fig.~\ref{fig:kink_th10_middle}.
The upper panel shows the birds eye view of $E(x,z)$.
The slightly bent wall arises near by the black hole.
The lower panel shows the contour plot of $E(x,z)$.
Each line corresponds to $E=0.1,0.2,0.3,0.4$.
The circle $\rho=1$ is the horizon.
The energy density distribution is almost homogeneous along the wall.
}
\label{fig:wall_th10_middle}
\end{figure}

\begin{figure}
\centerline{
\epsfxsize=14.0cm
\epsfbox{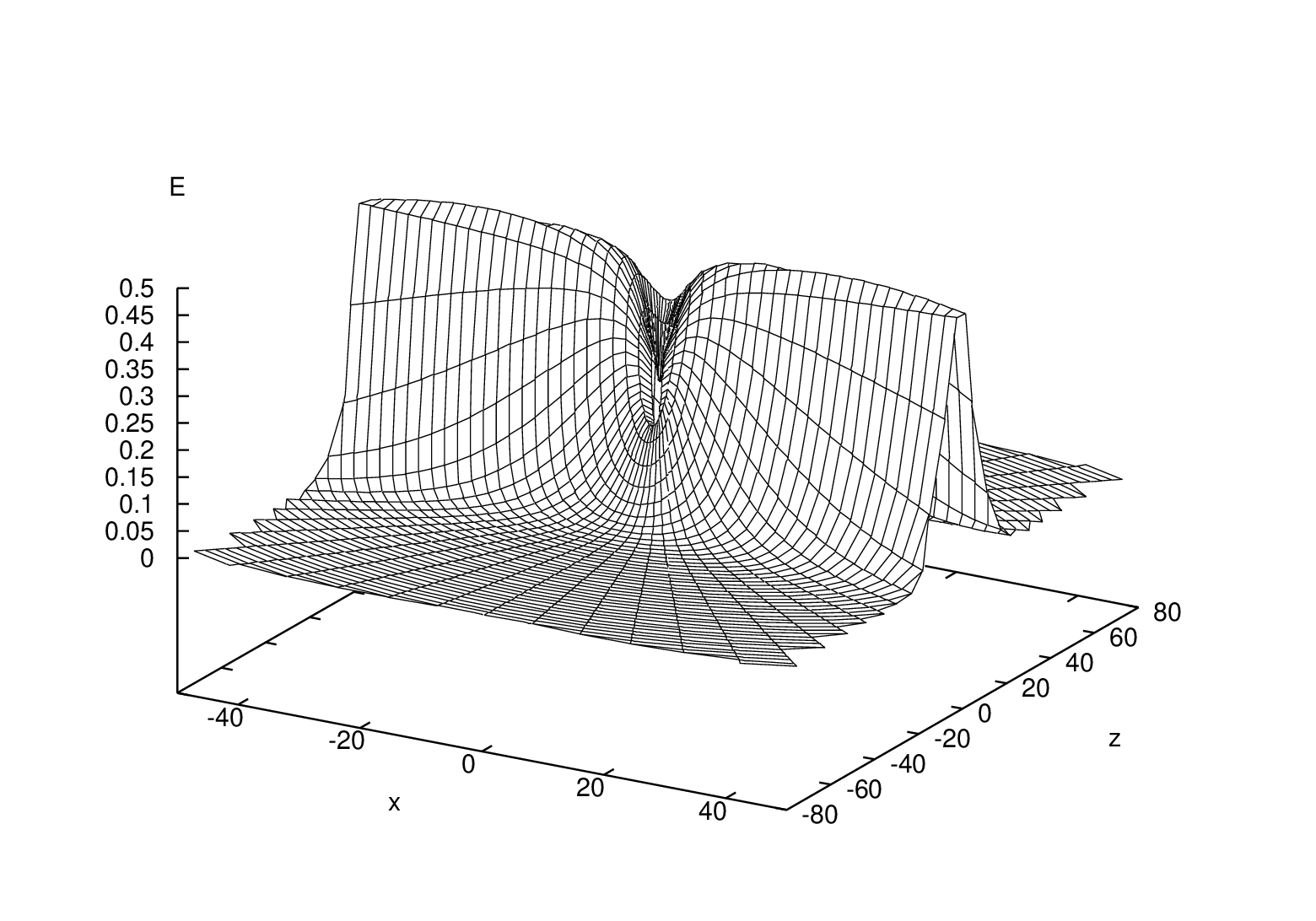}
}
\centerline{
\epsfxsize=14.0cm
\epsfbox{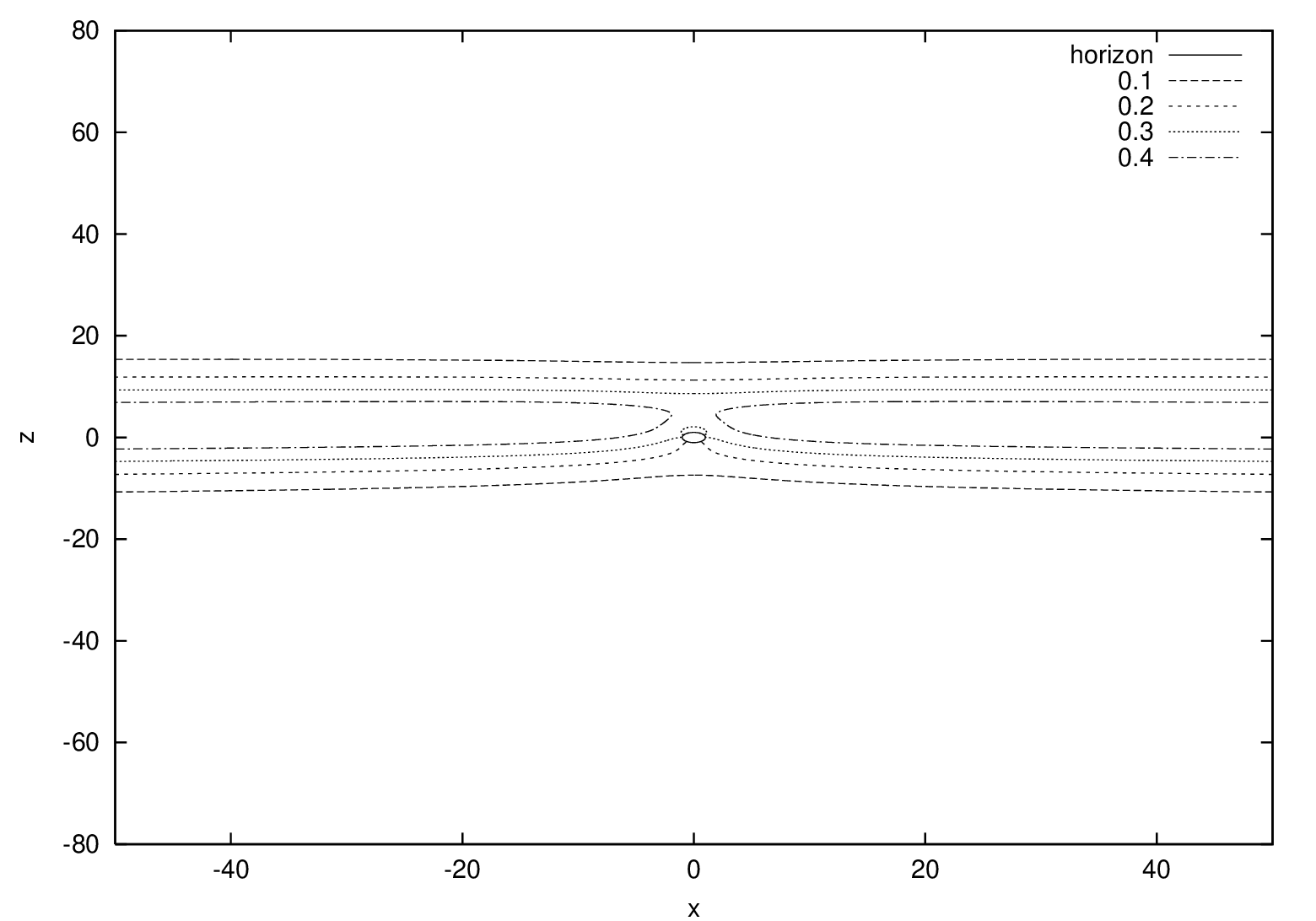}
}
\caption{(c) The energy density $E(x,z)$ of the wall solution in 
Fig.~\ref{fig:kink_th10_near}.
The upper panel shows the birds eye view of $E(x,z)$.
We see that the black hole is inside the thick wall.
The energy density reduces near the horizon.
The lower panel shows the contour plot of $E(x,z)$.
Each line corresponds to $E=0.1,0.2,0.3,0.4$.
The circle $\rho=1$ is the horizon.
The energy density distribution near the horizon is not homogeneous
along the wall, so that the Nambu--Goto approximation is not applicable
to this case.
}
\label{fig:wall_th10_near}
\end{figure}

\begin{figure}
\centerline{
\epsfxsize=14.0cm
\epsfbox{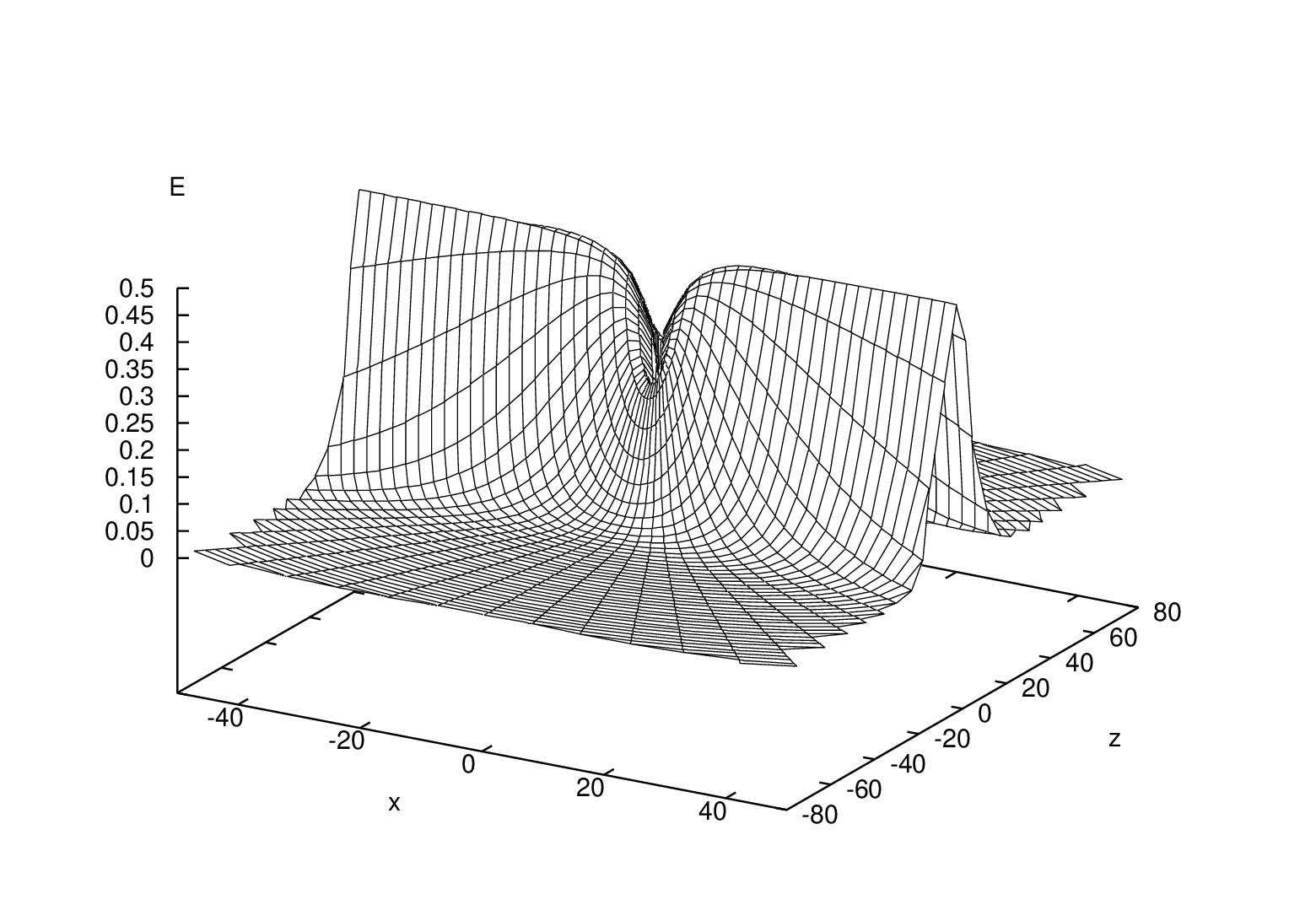}
}
\centerline{
\epsfxsize=14.0cm
\epsfbox{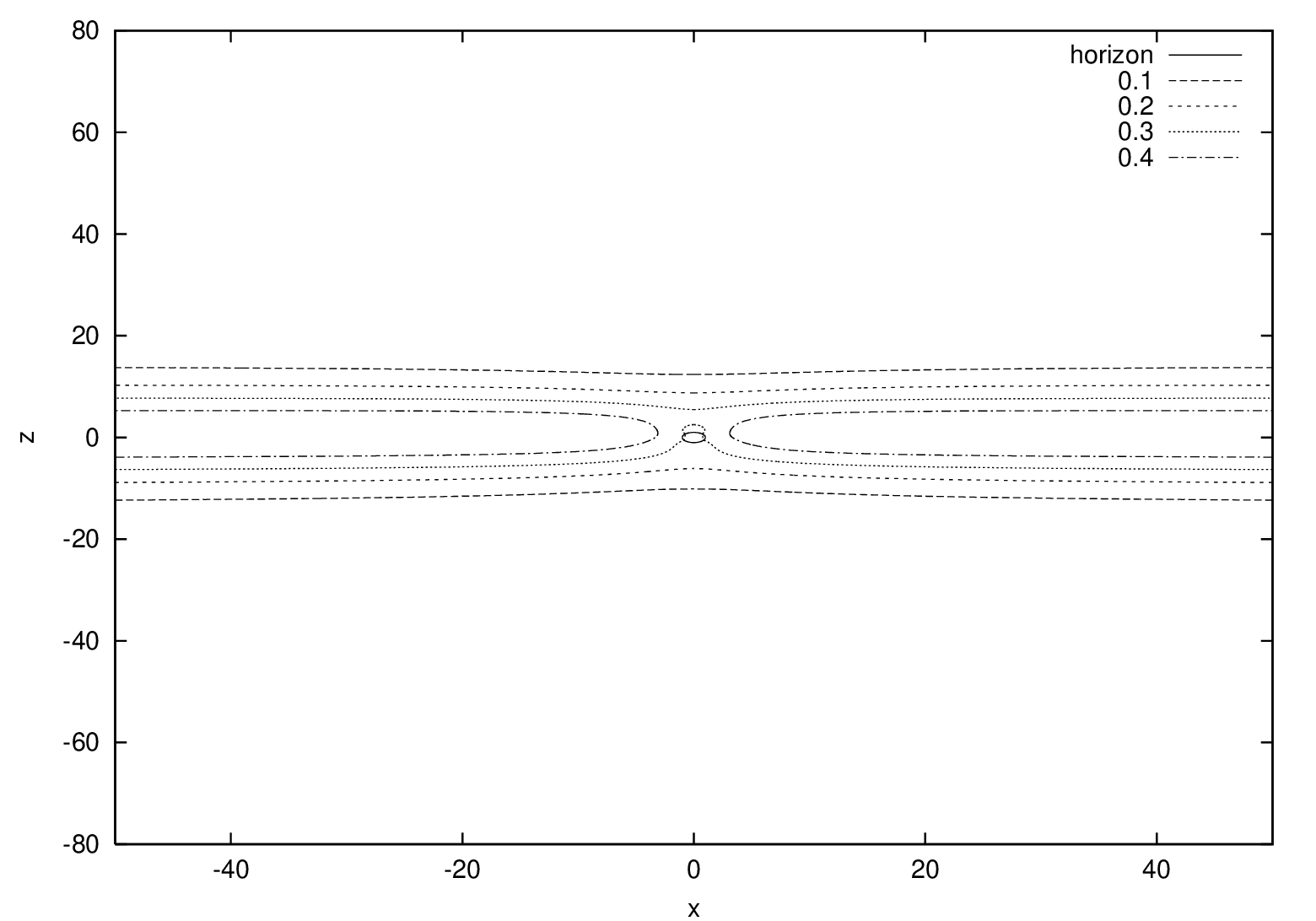}
}
\caption{(d) The energy density $E(x,z)$ of the wall solution in 
Fig.~\ref{fig:kink_th10_cross}.
The upper panel shows the birds eye view of $E(x,z)$.
We see that the black hole is inside the thick wall.
The graph has a similar shape to that of the wall located on 
the equatorial plane~\cite{Morisawa:2000qq}.
The lower panel shows the contour plot of $E(x,z)$.
Each line corresponds to $E=0.1,0.2,0.3,0.4$.
The circle $\rho=1$ is the horizon.
The energy density distribution near the horizon is not homogeneous
along the wall, so that the Nambu--Goto approximation is not applicable
to this case.
}
\label{fig:wall_th10_cross}
\end{figure}

We also show the energy density $E$ of the scalar field given by 
\begin{equation}
\label{eq:def_energy}
E \equiv
\frac{|T_t{}^t|}{\lambda\eta^4}
=
\frac{1}{2\epsilon^2}\left(\frac{\rho}{\rho+1}\right)^{4}
\left[\left(\frac{\partial\Phi}{\partial\rho}\right)^2
+\frac{1}{\rho^2}\left(\frac{\partial\Phi}{\partial\vartheta}\right)^2\right]
+U[\Phi],
\end{equation}
in Figs.~\ref{fig:wall_th10_far}--\ref{fig:wall_th10_cross}
corresponding to
Figs.~\ref{fig:kink_th10_far}--\ref{fig:kink_th10_cross},
respectively. 
In each figure,
the upper panel shows the birds eye view of $E(x,z)$,
and the lower panel shows the contour plot of $E(x,z)$.
Each line corresponds to $E=0.1,0.2,0.3,0.4$.
The solution (a) represents the almost flat domain wall with the
thickness $w\sim10$ arises far away from the black hole
(Fig.~\ref{fig:wall_th10_far}).
The solution (b) represents the slightly bent wall arises near by
the black hole
(Fig.~\ref{fig:wall_th10_middle}).
One can see from Figs.~\ref{fig:wall_th10_far} and \ref{fig:wall_th10_middle} 
that the walls can be set away from the black hole without a strong
disturbance by the black hole,
as expected from the asymptotic flatness of the background. 
In particular,
the energy density distribution is almost homogeneous along the wall.
For the solution (c) and~(d),
(Figs.~\ref{fig:wall_th10_near} and~\ref{fig:wall_th10_cross}),
we see that the energy density distributions near the horizon are
distorted and are not homogeneous along the wall, so that the thin
wall approximation is not applicable to this case.
This energy density distortion comes from the perpendicular
pressure as we shall see in the next section.

We shall comment on the numerical computation.  
The size of our integration domain is 250~times as large as 
the horizon radius ({\it i.e.} $\rho_{\rm max}=251$). 
The grid spacing in the $\rho$- and $\vartheta$-directions are 
$1$ ($\times$ horizon-radius) and $\pi/2^9$~(radian),
respectively.
After the relaxation has converged, we clip the region where $\rho \le 41$ 
in the neighborhood of the black hole, and take finer grid 
such that the grid spacing in the $\rho$-direction is $1/8$ 
($\times$ horizon-radius) in the clipped region.   
Then we restart the relaxation on this new finer grid.


\section{Comparison with Nambu--Goto membrane}
\label{sec:comparison}

It is non-trivial whether or not the core surfaces of the wall
configurations are actually well described by
the worldsheet of Nambu--Goto membranes.
If it is the case, one might be able to understand the qualitative
behavior of domain walls even in the thick wall case
by means of Nambu--Goto membranes, which are much easier to analyze
than scalar field configurations.

\begin{figure}
\centerline{
\epsfxsize=14.0cm
\epsfbox{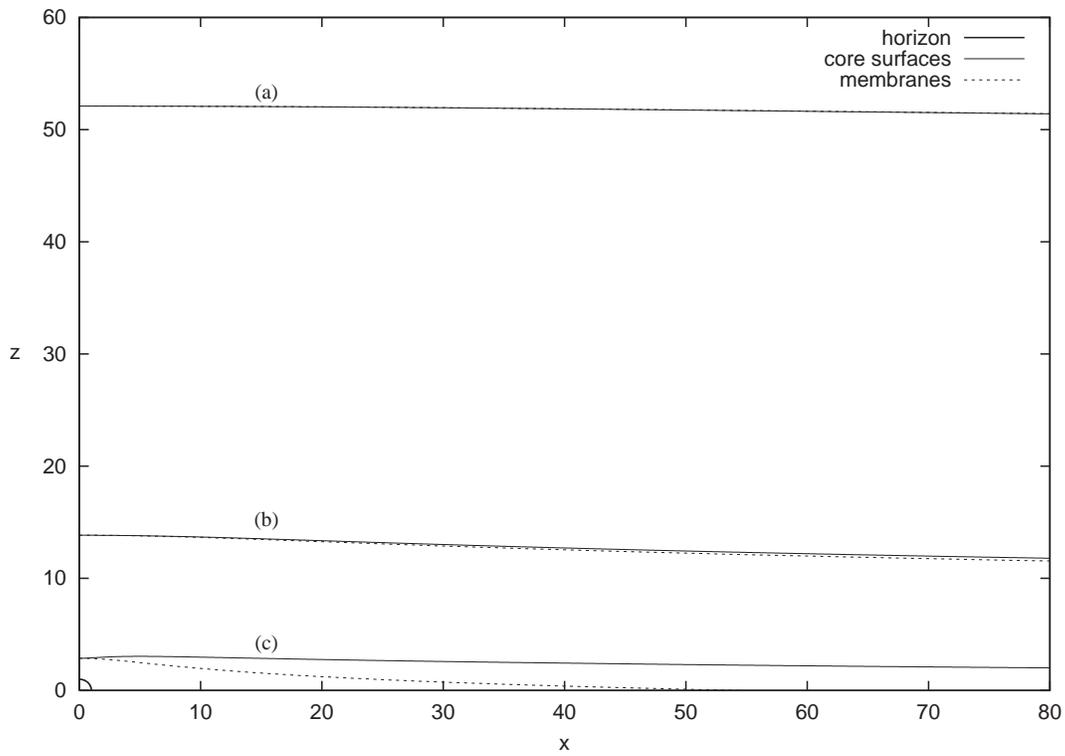}
}
\caption{We plot the core surfaces (where $\Phi=0$) of the three thick
wall solutions (a), (b) and (c).
For each core surface, a Nambu--Goto membrane tangent to the surface
at the point on the symmetry axis is plotted.
We see that the membranes well approximate the core surfaces for the
walls (a) and (b), but does not for (c).
The Nambu--Goto approximation in a black hole spacetime breaks down near
the horizon.
Here the coordinates $(x,z)$ are isotropic coordinates.
}
\label{fig:membrane-core}
\end{figure}

To compare the configuration of the core surfaces with the Nambu--Goto
membranes, we plot the core surfaces (where $\Phi=0$) of the three thick
wall solutions (a), (b) and (c)
in Fig.~\ref{fig:membrane-core}.
For each core surface, the Nambu--Goto membrane tangent to the surface
at the point on the symmetry axis is plotted.
We see that the membranes well approximate the core surfaces for the
walls (a) and (b), but does not for (c).
This shows that, when thick domain walls in a black hole spacetime are
concerned,
the approximate description of the core surface by means of
a Nambu--Goto membrane breaks down near the horizon.

Now we shall briefly discuss the reason why
the Nambu--Goto approximation breaks down,
examining the structure of
the energy-momentum tensor for the scalar field.
The energy-momentum tensor obtained from the Lagrangian~(\ref{eq:Lag}) is
\begin{equation}
T_{\mu\nu} = (\nabla_{\mu}\phi)(\nabla_{\nu}\phi)
-g_{\mu\nu} \left[
\frac{1}{2}(\nabla_{\sigma}\phi)(\nabla^{\sigma}\phi)+V
\right].
\end{equation}
Let us consider the following quantities defined by
\begin{eqnarray}
\lambda\eta^4 P_{\perp}
&\equiv& T_{\mu\nu}n^{\mu}n^{\nu}
= \frac{1}{2}(\nabla_{\sigma}\phi)(\nabla^{\sigma}\phi)-V,
\\
\lambda\eta^4 E
&=& \frac{1}{2}(\nabla_{\sigma}\phi)(\nabla^{\sigma}\phi)+ V,
\end{eqnarray}
where $n^{\mu}$ denotes the unit normal vector to the core surface,
given by $n^{\mu} = N\nabla^{\mu}\phi$ with
$N^{-2}=(\nabla_{\mu}\phi)(\nabla^{\mu}\phi)$, and $E$ is identical with
the energy density defined in Eq.~(\ref{eq:def_energy}).
Then, the energy-momentum tensor $T_{\mu\nu}$ can be expressed as
\begin{equation}
T_{\mu\nu} = - \lambda\eta^4 E(g_{\mu\nu}-n_{\mu}n_{\nu})
+ \lambda\eta^4 P_{\perp} n_{\mu}n_{\nu}.
\label{eq:energy-momentum}
\end{equation}
The first term of Eq.~(\ref{eq:energy-momentum}) looks just like
the energy-momentum tensor for a Nambu--Goto membrane,
once $\lambda\eta^4 E$ and $g_{\mu\nu}-n_{\mu}n_{\nu}$ are identified with
the tension and the induced metric, respectively.
The second term can be interpreted to describe the pressure
perpendicular to the wall and thus the deviation from the Nambu--Goto
membrane.
Non-vanishing $P_{\perp}$ means that
the gradient term and the potential term do not have the same
contribution to the energy density $E$, {\it i.e.},
$\lambda\eta^4 E\ne 2V$.

Next, let us derive the equation to decide the configuration of
the core surface.
The energy-momentum tensor~(\ref{eq:energy-momentum}) satisfies
the usual conservation law $\nabla_{\nu}T_{\mu}{}^{\nu}=0$.
Then, noticing that the trace of the extrinsic curvature $K_{\mu \nu}$
of the core surface is given by
\begin{equation}
\mbox{tr} K = \nabla_{\nu}n^{\nu},
\end{equation}
(in this convention, $\mbox{tr} K$ is positive
for outward normal $n$ of a sphere),
we can find that the normal component of the conservation law,
\begin{equation}
n^{\mu}\nabla_{\nu}T_{\mu}{}^{\nu}=0 ,
\end{equation}
reduces to the equation,
\begin{equation}
\mbox{tr} K = - \frac{n^{\mu}\partial_{\mu}P_{\perp}}{E+P_{\perp}}.
\label{eq:energy-conservation}
\end{equation}
The core surface obeys Eq.~(\ref{eq:energy-conservation})
while the Nambu--Goto membrane obeys $\mbox{tr}K=0$ as is well-known.

In order to make the meaning of the balance equation (20),
the spatial configuration of the domain wall is more useful.
Thus we rewrite Eq.~(20) by using the extrinsic curvature $\kappa_{ij}$
which specifies how the core surface is embedded in the constant time
hypersurface, not  in the whole spacetime.
The trace $\mbox{tr}\kappa$ of the extrinsic curvature $\kappa_{ij}$ is
related to $\mbox{tr}K$ as
\begin{equation}
\mbox{tr}K = \mbox{tr}\kappa + n^{\mu}\partial_{\mu}\ln\sqrt{-g_{tt}}.
\label{eq:trKtrK}
\end{equation}
The second term of Eq.~(\ref{eq:trKtrK}) can be rewritten by
the gradient of the Newton potential $\Phi_N$ in the weak field limit.
From Eqs.~(\ref{eq:energy-conservation})
and~(\ref{eq:trKtrK}), we obtain the following equation:
\begin{equation}
\mbox{tr}\kappa =
- n^{\mu}\partial_{\mu}\Phi_N
- \frac{n^{\mu}\partial_{\mu}P_{\perp}}{E+P_{\perp}}.
\label{eq:force_balance}
\end{equation}
The equation~(\ref{eq:force_balance})
requires the balance of the three forces:
the tension, the gravitational force, and the pressure gradient.

\begin{figure}
\centerline{
\epsfxsize=14.0cm
\epsfbox{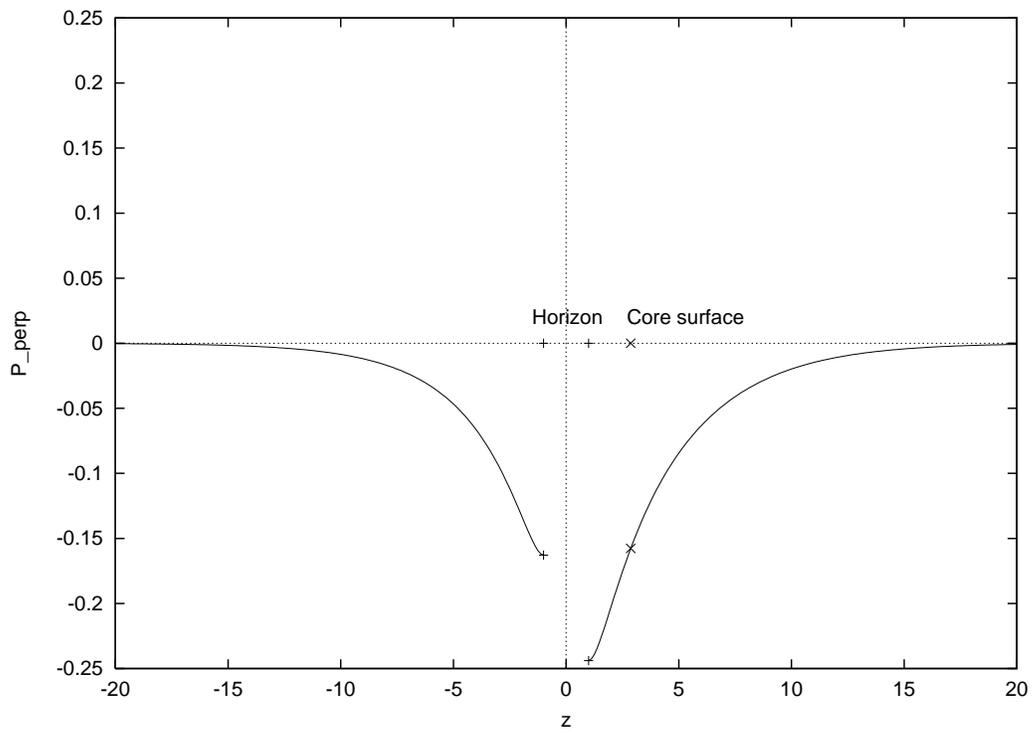}
}
\caption{The value of $P_{\perp}$ on the symmetry axis for the
wall (c) is plotted.
Two lines are split by the event horizon.
The position of the core surface is also shown.
The $P_{\perp}$ has negative value near the
horizon and asymptotes to zero.
}
\label{fig:P_perp}
\end{figure}

The distortion of the energy density near the horizon of the walls (c)
and (d) in the previous section implies the existence of $P_{\perp}$.
The $P_{\perp}$ for the wall (c) has negative value near the
horizon and asymptotes to zero as in Fig.~\ref{fig:P_perp}.
Then,
Eq.~(\ref{eq:force_balance}) implies that
the configuration of the core surface of (c) is
more convex in the direction of the horizon than
that of the corresponding Nambu--Goto membrane
as in Fig.~\ref{fig:membrane-core}.


\section{Summary and Discussion}     
\label{sec:summary}

We have numerically solved the equation of motion for a real scalar field 
with $\phi^4$ potential which have a discrete set of degenerate minima, 
in the Schwarzschild black hole background. 
We showed that there exist the static axi-symmetric field configurations 
which represent domain walls ten times as thick as the horizon radius
located around the black hole.
There are two types of the wall configurations; 
ones are far away from the black hole and the others are not.
As naturally expected from asymptotic flatness of the background
spacetime, the wall configurations in the former family are
similar to the walls in the flat spacetime and are well
approximated by the Nambu--Goto membranes.
For the wall solutions in the latter family,
we have obtained the wall solution whose core surface shows
different behavior from the Nambu--Goto membrane.
Then, we have shown the existence of  
the pressure gradient~$\partial_{n}P_{\perp}$ along 
the transverse direction to the core surface
of that solution.
The pressure~$P_{\perp}$ never has a place in the walls in the 
flat spacetime and Nambu--Goto membranes.

If the black hole is absent ({\it i.e.}, for the hyperbolic
tangent walls in the flat spacetime), the gradient term and the
potential term in the expression of $P_{\perp}$ cancel each other.
On the other hand, we impose the Neumann boundary condition on the 
symmetry axis and the horizon of the Schwarzschild black hole, and 
then the gradient term for static configurations vanishes at the
north pole and the south pole of the horizon.
Thus we may say that the gradient term cannot become large enough
to balance the potential term caused by the wall near these points,
and then the $P_{\perp}$ is negative there.
When the $P_{\perp}$ is negative near the horizon and asymptotes
to zero, Eq.~(\ref{eq:force_balance}) shows that
the configuration of the core surface in the constant time
hypersurface is more convex in the direction of the horizon than
that of the corresponding Nambu--Goto membrane.

One way of thinking about domain wall configurations we obtained here 
is in terms of gravitational scattering and capture of thick domain 
walls by a black hole. 
One might be able to think that a series of the wall solutions 
represents adiabatic capture process with infinitesimal velocity,
with the wall lying in the equatorial plane~\cite{Morisawa:2000qq}
being a possible final state.
This picture should be examined by a fully dynamical computation.

In the previous and present works, we have considered 
the Schwarzschild black hole as the background spacetime. 
One might expect that the scalar field would 
behave qualitatively in the same way as that in the Schwarzschild case, 
as far as static spherically symmetric black holes are considered 
as the background.  
However, the behavior of the scalar field near the event horizon 
indeed depends on what kind of a black hole one considers. 
It is amusing to note that a contrastive case to our present result 
is the case of extremal black holes. As is well-known, the behavior of 
gravity around an extremal black hole is quite different 
from the Schwarzschild black hole; an extremal black hole 
has a vanishing surface gravity hence the static observers along 
the extremal horizon are not accelerated. 
Emparan, Gregory, and Santos~\cite{Emparan:2000fn}
and~Rogatko~\cite{Rogatko:2001vv}
investigated domain walls intersecting charged and dilaton extremal 
black holes and showed that there are domain wall solutions whose 
contour lines wrap the event horizon with the scalar field remaining 
in the symmetric phase on the horizon. 
One can say in other words that extremal black holes expel the domain wall. 

In this work, we ignored the effect of the gravity of the wall
on the black hole spacetime. 
As mentioned in~\cite{Morisawa:2000qq}, this test wall assumption is 
valid as far as concerning very light domain walls with the symmetry 
breaking scale being much lower than the Planck scale. 
It has been shown~\cite{Bonjour:1999kz} that gravitating domain walls 
change the global geometry drastically;  
they makes spatial section compact. 
The equation of motion of a self gravitating thick domain wall
is perturbatively evaluated in~\cite{Bonjour:2000ca}.
It is fair to say that our solutions should be regarded 
to describe the local behavior of thick domain walls around a non-extremal 
black hole.

\section*{Acknowledgments}

We would like to thank Professors Takashi~Nakamura 
and Hideki~Ishihara for many useful suggestions and comments. 
This work is supported in part by the Japan Society 
for the Promotion of Science (A.I. and D.I.).  

\end{document}